\documentclass[twocolumn,amsmath,amssymb,pra,showpacs,floatfix]{revtex4-2}
\usepackage{graphicx,amsmath,amssymb,amsfonts,latexsym,dcolumn,bm,subfigure,float}
\usepackage{multirow}
\usepackage[usenames,dvipsnames]{color}
\usepackage{siunitx,braket,physics}

\begin{document}
\title{Casimir-Polder interactions of Rydberg atoms \\
with graphene-based van der Waals heterostructures}
\author{K. Wongcharoenbhorn$^1$, C. Koller$^2$, T.M. Fromhold$^1$, W. Li$^1$}
\affiliation{$^1$School of Physics and Astronomy, University of Nottingham, Nottingham NG7 2RD, UK}
\affiliation{$^2$University of Applied Sciences Wiener Neustadt, Johannes Gutenberg-Stra{\ss}e 3, 2700 Wiener Neustadt, Austria}
\date{\today}

\begin{abstract}
We investigate the thermal Casimir-Polder (CP) potential of \textsuperscript{87}Rb atoms
in Rydberg $n$S-states near single- and double-layer graphene. The dependence of the CP potential on parameters such as atom-surface distance, temperature, principal quantum number $n$ and graphene Fermi energy are explored. Through large scale numerical simulations, we show that, in the non-retarded regime, the CP potential is dominated by the
non-resonant and evanescent-wave terms which are monotonic, and that, in the retarded regime, the CP potential exhibits spatial oscillations. We identify that the most important contributions to the resonant component of the CP potential come from the $n$S-$n$P and $n$S-$(n-1)$P transitions. Scaling of the CP potential as a function of the principal quantum number and temperature is obtained. A heterostructure comprising hexagonal boron nitride layers sandwiched between two graphene layers is also studied. When the boron nitride layer is sufficiently thin, the CP potential can be weakened by changing the Fermi energy of the top graphene layer. Our study provides insights for understanding and controlling CP potentials experienced by Rydberg atoms near single and multi-layer graphene-based van der Waals heterostructures. 
\end{abstract}
\maketitle

\section{Introduction}
The development of ultracold-atom physics as platforms for chip-based matter wave manipulation \cite{Chen_Fan_2020}, high-accuracy time keeping systems \cite{liu_chen_2018, Ren_Li_2020}, quantum computing and simulation \cite{UKRev,EURev, Meinert_Christian_2020, Sebastian_Nithiwadee_2021} is an active research field. Understanding atom-surface interactions is essential for achieving near-surface atom trapping, as required for the operation of micro-fabricated atom chips. There is a substantial body of research on trapping ground-state atoms in metallic-wire-based atom chips \cite{Theodor_Bennett_2019, scheel2005, rekdal2004, folman2002, henkel1999, hinds_hughes_1999, jones2003} as well as on the interactions of ground-state atoms with various structures of metallic and perfect conductors \cite{Messina_Passante_2007, Vasile_Passante_2008, Alessandro_Passante_2019}. 
A conclusive review of atom-surface physics can be found in \cite{Laliotis_David_2021}.
%It is known that ground-state atoms have weaker and shorter range of interactions as well as shorter lifetimes compared to Rydberg atoms \cite{Jones_Marcassa_2017}, 
It is known that metallic-wire-based atom chips generate spatially rough trapping potential due to imperfections in the wires \cite{Sinuco_Fromhold_2011, Sinuco_Fromhold_2018, aigner_pietra_2008, japha_entin_2008, schumm_aspect_2005}, high Johnson-noise currents and strong Casimir-Polder (CP) attractive interactions between the atoms and the chip, causing, for example, tunnelling losses \cite{Henkel2003, Lin2004, Laliotis_Lu_2021}. In order to  enhance the functionality of such atom chips, different materials are needed for the current-carrying wires. Recent studies have shown that two-dimensional (2D) materials could offer desirable properties for overcoming the limitations of using metallic conductors as current-carrying wires \cite{Sinuco_Fromhold_2018, Wongcharoenbhorn_Crawford_2021}.\\
\\
There is also enduring interest in the technological applications of two-dimensional (2D) materials, including graphene, for display devices \cite{Ergoktas_Gokhan_2021}, flexible sensors \cite{Mufeng_Zheling_2020, kim_kim_2020, Carvalho_Kulyk_2021}, photo detectors \cite{Vaidotas_Simone_2020, Jacek_Mahmoud_2020, Jacek_Jacob_2020, Zhu_Hongxing_2021, AlAloul_Rasras_2021}, vertical field-effect transistors \cite{britnell2013, liu_liu_duan_2020}, and atom chips \cite{Wongcharoenbhorn_Crawford_2021}. New properties and methods of cooling or patterning 2D materials have been studied \cite{Ribeiro_Hugo_2016, Candussio_Durnev_2021, Tripathi_Lee_2021, ta_bachmatiuk_2020, Wei_Hauke_2021, Lin_Jingang_2021, Liu_Mohideen_2021, slizovskiy_Garcia_2021, greenaway_graphenes_2021}. As current-carrying wires in atom chips, graphene has desirable electronic properties: it has a very low density of electronic states, high carrier mobility and a linear band structure with zero band gap in the vicinity of the Dirac points \cite{mccann_electronic_2012}, which leads to Johnson noise and CP attraction far below those typically found for metallic conductors on bulk substrates \cite{Sinuco_Fromhold_2011, Fermani_Scheel_2007, Sinuco_Fromhold_2018, Wongcharoenbhorn_Crawford_2021}.
Importantly, smooth trapping potentials can be obtained using graphene. For example, 
graphene made from a helium-ion beam lithography technique
has edge roughness of order $\SI{5}{nm}$ \cite{aigner_pietra_2008}, while the surface roughness of graphene encapsulated in hexagonal boron nitride (hBN) is on the order of \SI{12}{pm} \cite{thomsen_gunst_2017}.
The coupling of atoms with graphene's surface plasmons might be tuneable via changing the Fermi energy of graphene as the plasmon frequency is proportional to the fourth root of its electronic density \cite{ju_graphene_2011, yan_damping_2013, cui_graphene_2021, Haakh_Passante_2014, Xiaodong_Suhail_2021}. Previous studies have shown that this could be used to tailor the CP interactions of atoms trapped near graphene \cite{Cysne_Kort_2014,Liu_Mostepanenko_Mohideen_2021,Henkel_Klimchitskaya_2018,Chaichian_Tureanu_2012, churkin2010, Bordag_Mostepanenko_2006, Cysne_Peres_2016, Nichols_Valeri_2016, Klimchitskaya_Vladimir_2020, Kaur_Kaur_2014, Khusnutdinov_2018}.\\
\begin{figure}[ht]
\centering
\includegraphics[width = 0.8\linewidth]{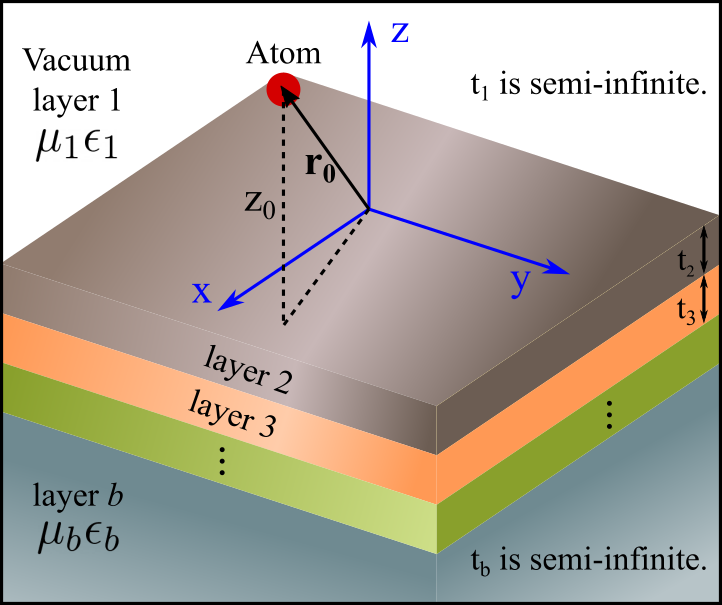}
\caption{Schematic diagram of an $b$-layer system, where each layer is designated by index $q = 1, 2,..., b$ and is characterized by thickness, $t_{q}$, permeability, $\mu_{q}$, and permittivity, $\epsilon_{q}$. An atom (red circle) is located at $\mathbf{r}_{0} = (x_{0}, y_{0}, z_{0})$ in layer 1 which is vacuum. Note that, in the Green's-function calculations, all layers are infinitely extended in the $x-y$ plane, and that $t_{1}$ and $t_{b}$ are also infinite.}
\label{fig:Planar_structure}
\end{figure}
\\
In recent years, the study of interactions between Rydberg atoms and surfaces has attracted much attention.  Such studies could help us understand the atom-surface interaction for highly excited atomic states, and open new quantum technological applications \cite{Nikola_Charles_2018}, for example by integrating Rydberg atoms with condensed matter quantum materials like graphene. Rydberg atoms are highly excited atoms with large principal quantum number, i.e. $n \gg 1$. The size of Rydberg atoms is proportional to $n^{2}$, and can be a micron when $n \sim 90$, resulting in weakly bound valence electrons and high electric polarizability, which in turn causes strong interactions with nearby surfaces \cite{kohlhoff_interaction_2016}. The lifetime associated with spontaneous emission, i.e. at $\SI{0}{K}$, is proportional to $n^{3}$. The long lifetime of Rydberg states allows us to exploit Rydberg atoms for quantum computing and simulation \cite{Hermann_2014, Wu_Liang_2021}.  Atomic transition frequencies between adjacent Rydberg states are typically in the microwave-terahertz regions, which are in the same window as thermal energies at room temperature; Rydberg atoms interact resonantly with thermal photons, leading to enhanced CP interactions compared to ground-state atoms \cite{Ellingsen_Scheel_2010}. Strong coupling between Rydberg atoms and surface plasmon polaritons or surface phonon polaritons have also been studied both experimentally and theoretically \cite{T_rm__2014, Kubler_Booth_2013, Ribeiro_Buhmann_Scheel_2013, Sheng_Chao_2016}. 
There have been previous studies of the CP interactions of excited two-level or realistic Rydberg atoms near mirrors, metallic and dielectric surfaces, and metamaterials \cite{Barton_1974, Power_Thirunamachandran_1982, Barton_1988, Meschede_Hinds_1990, Hinds_Sandog_1991, Bambini_Edward_1992, Fichet_Ducloy_1995, gorza_ducloy_2001, Bushev_Rainer_2004, Buhmann_Trung_2004, gorza_ducloy_2006, Buhmann_Gunnar_2008, Schiefele_Henkel_2010, crosse_buhmann_scheel_2010, Antezza_Salvatore_2014, Ribeiro__Scheel_2015, Armata_Passante_2016, Carvalho_Laliotis_2018, Yang_Zhang_2019, Wei_Yang_2019, Buhmann_Singer_2021}, as well as investigations of the CP interactions of a laser-driven atom and a surface \cite{Sebastian_Stefan_Yoshi_2018}. The CP interactions between excited molecules or ions and metallic or dielectric bodies have also been studied \cite{Wilson_Dorner_2003, Bushev_Rainer_2004, Barcellona_Yoshi_2016}. However, CP interactions between Rydberg atoms and 2D nano-structures such as graphene have not been fully explored. To the best of our knowledge, the Rydberg atom-graphene interaction has only been studied in Ref. \cite{Ribeiro_Scheel_2013}, which focuses on the zero-temperature limit and shows that graphene can shield the CP force emanating from a metallic substrate when graphene-substrate distances are larger than $\SI{4}{\mu}$m.\\
\\
Integrating Rydberg atoms with graphene could lead to promising and novel quantum devices \cite{Muller_Henkel_2011, wade_weatherill_2016, saffman_2016}.
In this work, we focus on investigating the CP interactions of a \textsuperscript{87}Rb atom in Rydberg $n$S states with graphene. Two notable features, which are not found in ground-state atoms in the zero-temperature limit, emerge: stimulated atomic transitions due to thermal photons give rise to resonant interactions, which can be further distinguished into attractive and repulsive potentials, and the large sizes of Rydberg atoms give rise to quadrupole interactions at short atom-surface distances.\\
\\
The paper is organized as follows. In Sec. \ref{sec:Casimir-Polder Potential near Planar Structures}, we provide the quantum field theoretical description of the CP potential for atoms in Rydberg states and a brief description of Rydberg atoms. In Sec. \ref{sec:Different material models}, we first compare the CP potential of a \textsuperscript{87}Rb atom in Rydberg states near a suspended single layer of graphene whose conductivity is described by the Kubo model, and a $1$-$\mu$m-thick gold sheet whose optical properties are described by the Drude model. We also consider another model of graphene, with the results presented, which includes non-local effects such as lattice defects. In Sec. \ref{sec:Characteristics of the CP potential}, we present detailed calculations of the CP potential of single-layer graphene, which allow us to see various characteristics of Rydberg atoms near graphene that are different from those of ground-state atoms, especially effects arising from resonant interactions. In Sec. \ref{sec:Fitted empirical functions}, we find simple fitted empirical functions, which capture the scaling of the CP potential with principal quantum number and the temperature dependence of the interaction. Finally, in Sec. \ref{sec:Double-Layer Graphene Heterostructures}, we extend our study into graphene-based van der Waals heterostructures comprising two layers of graphene separated by air or hexagonal boron-nitride (hBN). This allows us to change the spacing between the graphene layers and change their chemical potentials in order to tune the interactions between the trapped atoms and the heterostructures. We conclude in Sec. \ref{sec:conclusion}. 

% \section{Rydberg Atoms}
% \label{sec:Rydberg Atoms}
% An alkali-metal atom in a Rydberg state consists of a closed positive-ion core and a highly-excited valence electron moving in a large orbit under the influence of an almost hydrogenic Coulomb field generated by the core.

% \begin{figure}[ht]
% \includegraphics[width = \linewidth]{Radial_Wavefunction.png}
% \caption{}
% \label{fig:Radial_Wavefunction}
% \end{figure}

\section{Casimir-Polder Potential near Planar Structures}
\label{sec:Casimir-Polder Potential near Planar Structures}
In this section, we will present a theoretical description of the CP energy shift of highly excited Rydberg atoms near planar layered structures. Considering only ultracold atoms in Bose-Einstein condensates allows us to disregard the velocity-dependent CP potential since the atoms are not moving at a relativistic speed \cite{Scheel_Buhmann_2009}. Further details of the formalism used can be found in, for example, Refs. \cite{crosse_buhmann_scheel_2010, scheel2008, buhmann_ii, buhmann_thermal}.
The CP potential arises from the coupling between an atom and the surrounding body-modified electromagnetic radiation, described by the coupling Hamiltonian $\hat{H}_{\textrm{AF}}$. The interaction Hamiltonian in the case of a Rydberg atom can be split into dipole interactions (first term) and quadrupole interactions (second term): 
\begin{equation}
\hat{H}_{\textrm{AF}} = -\hat{\textbf{d}}\cdot\hat{\textbf{E}}(\mathbf{r}_{0}) - \hat{\textbf{Q}}\bullet\big[\boldsymbol{\nabla}\otimes\hat{\textbf{E}}(\mathbf{r}_{0})\big] ,
\label{eq:Hamiltonian}
\end{equation}
where $\hat{\textbf{d}}$ and $\hat{\textbf{Q}}$ are the atomic dipole moment and quadrupole moment operators, respectively, with $\hat{\textbf{E}}(\mathbf{r}_{0})$ being the electromagnetic field at the position of the atom, $\mathbf{r}_{0}$, $\bullet$ denotes the Frobenius inner product and $\otimes$ tensor product.
The energy-level shift up to the second order, for an atom in state $\ket{u}$ and the body-modified electromagnetic field in state $\ket{v}$, is \cite{crosse_buhmann_scheel_2010}
\begin{equation}
\delta E_{u} = \bra{u,v}\hat{H}_{\textrm{AF}}\ket{u,v} + \sum_{u^{\prime}, v^{\prime} \neq u,v}\frac{\abs{\bra{u,v}\hat{H}_{\textrm{AF}}\ket{u^{\prime},v^{\prime}}}^{2}}{E_{u, v} - E_{u^{\prime}, v^{\prime}}},
\label{eq:deltaE}
\end{equation}
where $E_{u, v}$ are the unperturbed energy eigenvalues of the atom-field system: $u$ and $v$ are the quantum numbers of the Rydberg state and the photon field, respectively. This energy shift $\delta E_{u}$ can be expressed in terms of the Green's tensor, which  can be decomposed into bulk and scattering parts. It follows that the energy shift can be split into two components: the position-independent self-energy (associated with the bulk part of the Green's tensor) similar to the Lamb shift and the position-dependent component (associated with the scattering part of the Green's tensor), which is, namely, the CP potential.\\

For an atom, at position $\mathbf{r}_{0} = (x_{0}, y_{0}, z_{0})$ in the planar system shown in Fig. \ref{fig:Planar_structure}, in an incoherent superposition of internal-energy eigenstates $\ket{u}$ (specified by the principal quantum number $n$, the orbital angular momentum quantum number $l$, the total angular momentum quantum number $j$, and the $z$-component of the total angular momentum quantum number $m$) with probabilities $p_{u}$ as described by a density matrix
\begin{equation}
\hat{\sigma} = \sum_{u} p_{u}\ket{u}\bra{u},
\label{eq:densitymatrixofatom}
\end{equation}
the total thermal CP potential at environment temperature $T$ can be written as \cite{scheel2008, buhmann_thermal}
\begin{equation}
\begin{split}
U_{\mathrm{CP}}(\mathbf{r}_{0}) &= \sum_{u}p_{u}U_{u}(\mathbf{r}_{0}),\\
U_{u}(\mathbf{r}_{0}) &= U_{u}^{\mathrm{nres}}(\mathbf{r}_{0}) + U_{u}^{\mathrm{res}}(\mathbf{r}_{0}).
\end{split}
\label{eq:nres_res}
\end{equation}
Here, $U_{u}(\mathbf{r}_{0})$ is the position-dependent part of $\delta E_{u}$,  $U_{u}^{\mathrm{nres}}(\mathbf{r}_{0})$ is the non-resonant potential due to virtual photons and $U_{u}^{\mathrm{res}}(\mathbf{r}_{0})$ is the resonant potential due to real thermal photons.\\

For the potentials due to dipole interactions, the non-resonant term is given by \cite{buhmann_ii}
\begin{equation}
U_{u}^{\textrm{nres}}(\mathbf{r}_{0}) = \mu_{0}k_{B}T\sum_{a=0}^{\infty}{}^{'}\xi_{j}^{2}\big[\boldsymbol{\alpha}_{u}(i\xi_{a})\bullet\mathbf{G}^{(\mathrm{s})}(\mathbf{r}_{0}, \mathbf{r}_{0}, i\xi_{j})\big],
\label{eq:U_CP^dip_nres}
\end{equation}
whereas, the resonant term is given by

\begin{multline}
U_{u}^{\textrm{res}}(\mathbf{r}_{0}) = -\mu_{0}\sum_{k<u}[N(\omega_{uk}) + 1]\omega_{uk}^{2}\\
\times\big(\textbf{d}_{uk}\otimes\textbf{d}_{ku}\big)\bullet\textrm{Re}\mathbf{G}^{(\mathrm{s})}(\mathbf{r}_{0}, \mathbf{r}_{0}, \omega_{uk})\\
+\mu_{0}\sum_{k>u}N(\omega_{ku}) \omega_{ku}^{2}\big(\textbf{d}_{uk}\otimes\textbf{d}_{ku}\big)\bullet\textrm{Re}\mathbf{G}^{(\mathrm{s})}(\mathbf{r}_{0}, \mathbf{r}_{0}, \omega_{ku}),
\label{eq:U_CP^dip_res}
\end{multline}
where $\mu_{0}$ is the permeability of free space, $k_{B}$ the Boltzmann constant, $\mathbf{G}^{(\mathrm{s})}$ is the scattering Green's tensor (see Appendix \ref{appendix:Green's tensor} for details), and $\textbf{d}_{uk} = \bra{u}\hat{\textbf{d}}\ket{k}$ is the dipole matrix element.
The atomic dipole polarizability as a function of radiation frequency, $\omega$, is defined as
\begin{equation}
\boldsymbol{\alpha}_{u}(\omega) = \lim_{\epsilon \to 0+}\frac{1}{\hbar}\sum_{k \neq u}\Bigg(\frac{\mathbf{d}_{uk}\otimes\mathbf{d}_{ku}}{\omega_{ku} - \omega - i\epsilon} + \frac{\mathbf{d}_{ku}\otimes\mathbf{d}_{uk}}{\omega_{ku} + \omega + i\epsilon}\Bigg),
\label{eq:dipole_polarizability}
\end{equation}
with $\omega_{ku} = (E_{k} - E_{u})/\hbar$ denoting the atomic transition angular frequencies. The purely imaginary frequencies $\xi_{j} = 2\pi k_{\mathrm{B}}Ta/\hbar$, $a = 0, 1, 2,...$ are the Matsubara frequencies and $N(\omega) = 1/[\mathrm{e}^{\hbar\omega/(k_{\mathrm{B}}T)} - 1]$ is the average thermal photon number in accordance with Bose-Einstein statistics. Following the separation of the scattering Green's tensor into propagating-wave and evanescent-wave components, the resonant term of the CP potential can also be split into two components as well.\\

In our calculations, the binding energy of the Rydberg series is given by $E_{n,l,j} = -\mathrm{Ry}/n^{*2}$ \cite{Gallagher_1988}, where $\mathrm{Ry}$ is the Rydberg energy and $n^{*} = n - \delta_{n,l,j}$. Here, $\delta_{n,l,j} = \delta_{0} + {\delta_{2}/(n - \delta_{2})^{2}}$ is the quantum defect \cite{Seaton_1983} whose values for \textsuperscript{87}Rb are tabulated in Table \ref{table:Quantum Defect}.

The electron wavefunction at position $\mathbf{r}$ with respect to the ion core, $\psi(\mathbf{r}) = \psi_{n,l,j,m}(\mathbf{r})$, for the valence electron is described by the Schr\"{o}dinger equation
\begin{equation}
\Big[-\frac{\hbar^{2}}{2\mu}\nabla^{2} - \frac{e^{2}}{4\pi\epsilon_{0}r}\Big]\psi(\mathbf{r}) = [E_{n,l,j} + V(\mathbf{r})]\psi(\mathbf{r}),
\label{eq:Schrodinger}
\end{equation}
where $e$ is the electronic charge, $\mu = Mm_{e}/(M+m_{e})$ is the reduced mass with $M$ being the mass of the atom, $m_{e}$ is the electronic mass, $r = \abs{\mathbf{r}}$ and $V(\mathbf{r})$ is a model potential as given in Ref. \cite{ARC_Robertson_2021}, which accounts for the finite size of the core at short range. Eq. \eqref{eq:Schrodinger} can be separated into angular and radial equations which yield, in standard notation, spherical harmonics $Y_{l,m_{l}}(\theta,\phi)$ and radial wavefunctions $R_{n,l,j}(r)$ as solutions, respectively. In this work, the radial wavefunctions are calculated numerically using the tool provided in Ref. \cite{ARC_Robertson_2021} through Numerov's method \cite{Numerov_1927}.

\begin{table}
\begin{tabular}{|c|c|c|} 
 \hline
\phantom{0}\textbf{State}\phantom{0} & \phantom{00000}$\delta_{0}$\phantom{00000} &
\phantom{00000}$\delta_{2}$\phantom{00000}\\ 
\hline
$n\mathrm{S}_{1/2}$ & 3.1311804 & 0.1784\\ 
\hline
$n\mathrm{P}_{1/2}$ & 2.6548849 & 0.2900\\ 
\hline
$n\mathrm{P}_{3/2}$ & 2.6416737 & 0.2950\\ 
\hline
$n\mathrm{D}_{3/2}$ & 1.34809171 & -0.60286\\ 
\hline
$n\mathrm{D}_{5/2}$ & 1.34646572 & -0.59600\\ 
\hline
\end{tabular}
\caption{Quantum defects for S, P and D states of Rb atoms  \cite{PhysRevA.67.052502}.}
\label{table:Quantum Defect}
\end{table}

% To ease our calculations, we will neglect the quadrupole interactions since it can be seen from \cite{crosse_buhmann_scheel_2010} that their contributions are small even at short atom-surface distances compared to those from the dipole ones; they will become weaker as the separation increases.

\section{Comparisons between material models} \label{sec:Different material models}
In this section, we will investigate how different conductivity models affect the CP potential. For graphene, we take the Fermi energy and electron relaxation rate of graphene to be $E_{F} = \SI{0.1}{eV}$ and $\gamma = \SI{4}{THz}$, respectively, corresponding to typical values found both theoretically \cite{gric_2019,andryieuski_lavrinenko_2013,Goncalves2016,amorim_2017} and in experiments \cite{ju_graphene_2011}, unless otherwise explicitly stated.

\subsection{Graphene vs Gold}
\begin{figure}[ht]
\includegraphics[width = \linewidth]{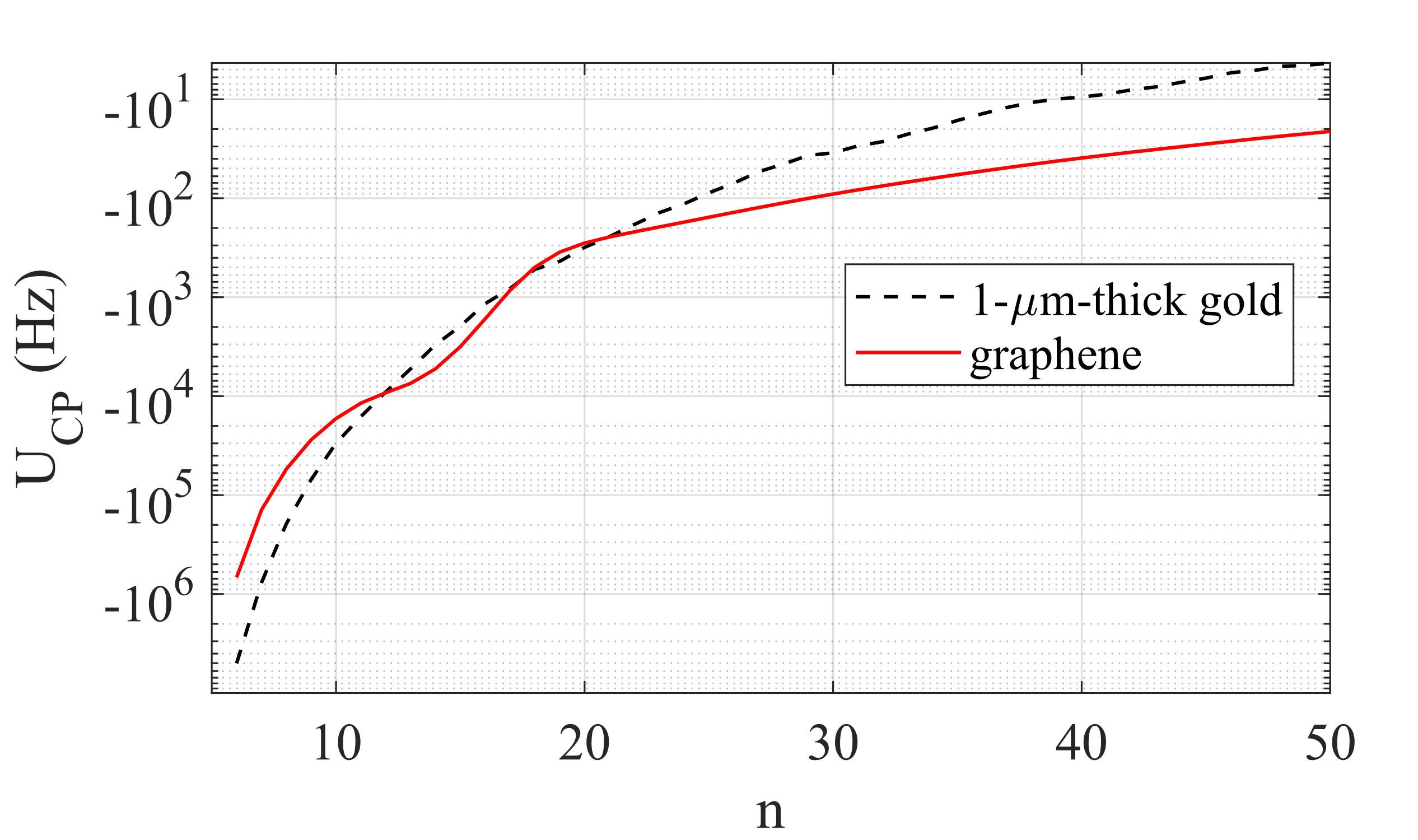}
\caption{CP potential energy calculated versus principal quantum numbers for an $^{87}$Rb atom near a $1$-$\mu$m-thick gold sheet (dashed curve) or a suspended single layer of graphene (solid curve). For each $n$, the atom-surface separation is taken to be $1/10$ of the wavelengths of the $n\mathrm{S}$-$(n-1)\mathrm{P}_{1/2}$ transitions. This is to ensure that the calculations are in the non-retarded regime. For $n < 12$, the CP potential for graphene is weaker than that for gold. For higher $n$, the relative strengths reverse until $n > 20$, when the potential of graphene again becomes stronger than that of gold.}
\label{fig:6_50SP_gold_Gr_vs_n}
\end{figure}
We first consider a free-standing graphene monolayer, which, from Fig. \ref{fig:Planar_structure}, can be modeled as a two-layer system, in which the monolayer graphene is located at the interface between layer 1 and layer 2. Its conductivity is modeled by the Kubo formula and is not a function of wavevector of impinging radiations (see Appendix \ref{appendix:Graphene's Optical Properties} for the description of the model). Note that throughout this paper, we will refer to the distance between an atomic core and a surface as the atom-surface distance. In Fig. \ref{fig:6_50SP_gold_Gr_vs_n}, we consider the CP potential in the non-retarded regime by choosing the atom-surface distance to be $1/10$ of the wavelength of the corresponding $n\mathrm{S}$-$(n-1)\mathrm{P}_{1/2}$ transition for each individual principal quantum number between $6$ and $50$. The atom-surface distances therefore range from $\SI{136}{nm}$ for $n = 6$ to $\SI{0.881}{mm}$ for $n = 50$. For comparison, we also show the CP potential for an atom near a $1$-$\mu$m-thick free-standing gold sheet in the same figure. Generally, the CP potential of ground-state atoms near graphene is weaker than for those near a thick metallic sheet \cite{Wongcharoenbhorn_Crawford_2021}. However, this is not always the case for Rydberg atoms. We can see that when $n < 12$, the CP potential for graphene is weaker than that for gold. The situation changes for $n \geq 21$: the CP potential for graphene becomes stronger than that for gold.\\

% \begin{figure}[htbp]
% \includegraphics[width = \linewidth]{U_CP_gold_graphene_30S.png}
% \caption{The CP potentials between a $1$-$\mu$m-thick gold sheet (black solid curve) and a suspended single layer of graphene (red dashed curve) for a \textsuperscript{87}Rb atom in the 30S state at (a) $T = \SI{10}{K}$ and (b) $T = \SI{300}{K}$. The CP potential for graphene is slightly stronger than that for gold at low temperature, which is not the case for ground-state atoms. At room temperature, the CP potential for graphene is weaker than that for gold.}
% \label{fig:U_CP_gold_graphene_30S}
% \end{figure}

% \begin{figure}[ht]
% \includegraphics[width = \linewidth]{20_40_gold_vs_Gr_2micron.png}
% \caption{The CP potential between a $1$-$\mu$m-thick gold sheet (black solid curve) and a suspended single layer of graphene (red dashed curve) for a \textsuperscript{87}Rb atom at $z_{0} = \SI{2}{\mu}$m for (a) $T = \SI{10}{K}$ and (b) $T = \SI{300}{K}$. The CP potential for graphene is slightly stronger than that for gold at $\SI{10}{K}$ for all of the principal quantum numbers considered here. At $T = \SI{300}{K}$, the CP potential for graphene gradually changes from being weaker than gold at $n = 20$ to being stronger towards $n = 40$.}
% \label{fig:20_40_gold_vs_Gr_2micron}
% \end{figure}

\subsection{Two models for monolayer graphene conductivity} 
\label{sec:different models}
\begin{figure}[ht]
\includegraphics[width = 0.8\linewidth]{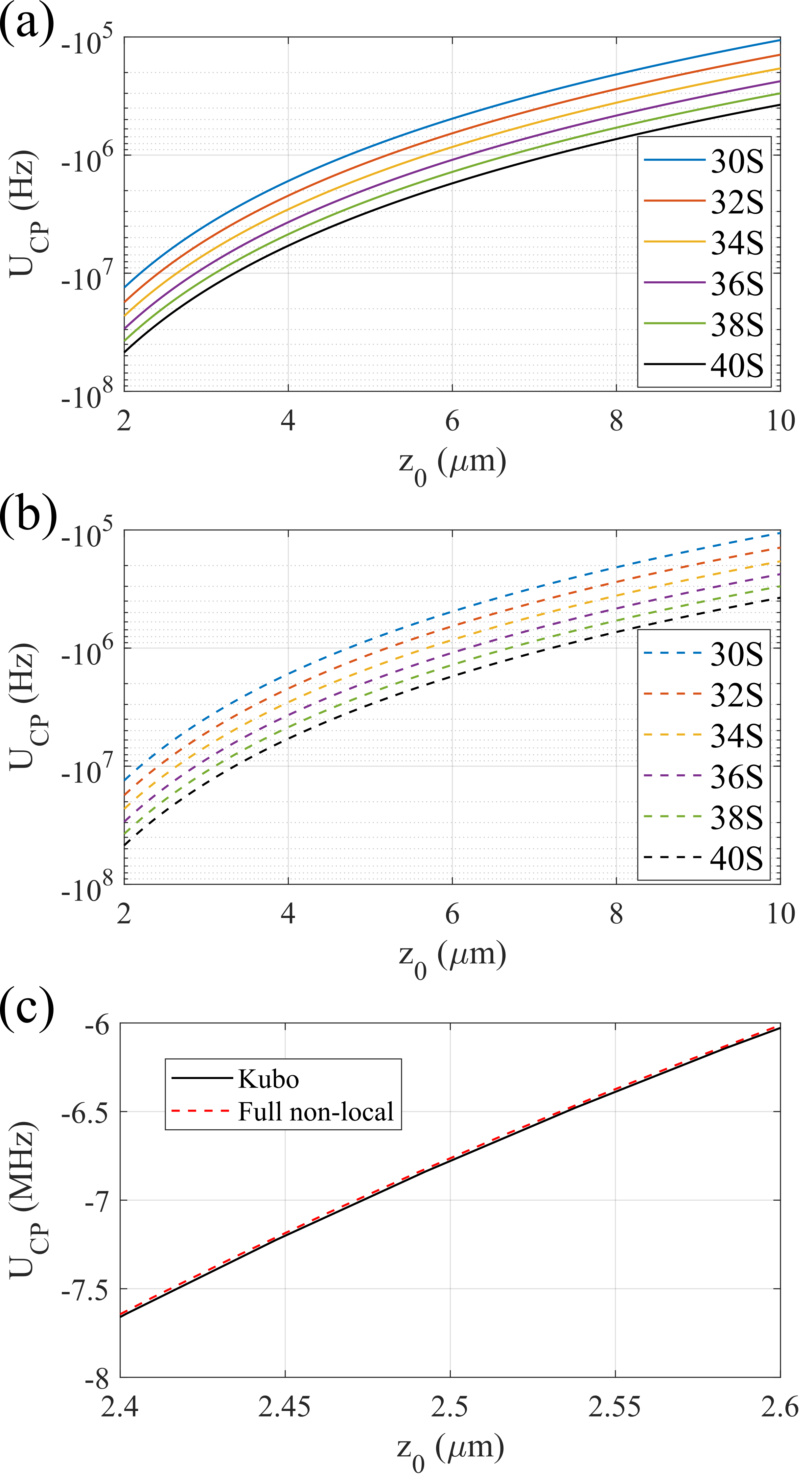}
\caption{CP potentials calculated versus the distance of an $^{87}$Rb atom from a graphene monolayer for two conductivity models of graphene: (a) Kubo conductivity, (b) full non-local conductivity. (c) shows the CP potential calculated for the 30S state at $T = \SI{10}{K}$ from the full non-local model (dashed curve) and Kubo model (solid curve).}
\label{fig:CompareTwoModels_5}
\end{figure}

Another conductivity model is the full non-local model, which includes non-local effects such as lattice defects. In this model, the conductivity is a function of both frequency and wavevector of impinging radiations; more details can be found in Appendix \ref{appendix:Graphene's Optical Properties} and in, for example, Ref. \cite{Goncalves2016}. The comparison of the CP potentials for multiple Rydberg states in the linear-log scale between the Kubo conductivity and the full non-local conductivity is shown in panels (a) and (b) of Fig. \ref{fig:CompareTwoModels_5}, and the CP potential for the 30S state is shown in (c). We can see that the potentials from the full non-local model are slightly weaker than those from the local conductivity model. The two models give consistent results in terms of magnitudes and spatial dependence. Due to the excellent agreement between the two models, we will utilize the Kubo model for shorter computational times in the following sections, unless otherwise explicitly stated. 

\section{Characteristics of the CP potential}
\label{sec:Characteristics of the CP potential}
In this section, we will explore the dependence of the CP potential on atom-surface distances, temperature and Rydberg states. Since the resonant part of the CP potential in Rydberg atoms is enhanced by thermal photons and is significantly larger than in ground-state atoms, we will also investigate in detail the contributions from individual atomic transitions. This will provide insights into the characteristics of the CP potential for Rydberg atoms.

\subsection{Characteristic quantities}
\label{sec:Limiting Cases}
\begin{table}[H]
\begin{tabular}{ |c|c|c|c| } 
\hline
\textbf{Limit} & \textbf{Condition} & $z$ & $T$ \\ 
 \hline
 Retarded & $z_{0}\omega_{-}/c \gg 1$ & $z_{\omega} \ll z_{0}$ & $T_{z} \ll T_{\omega}$ \\ 
 \hline
 Non-retarded & $z_{0}\omega_{+}/c \ll 1$ & $z_{0} \ll z_{\omega}$ & $T_{\omega} \ll T_{z}$ \\ 
 \hline
  Spectroscopic low-T & $k_{B}T \ll \hbar\omega_{-}$ & $z_{\omega} \ll z_{T}$ & $T \ll T_{\omega}$ \\ 
 \hline
  Spectroscopic high-T  & $k_{B}T \gg \hbar\omega_{+}$ & $z_{T} \ll z_{\omega}$ & $T_{\omega} \ll T$ \\ 
 \hline
  Geometric low-T  & $k_{B}T \ll \hbar c/z_{0}$ & $z_{0} \ll z_{T}$ & $T \ll T_{z}$ \\ 
 \hline
  Geometric high-T  & $k_{B}T \gg \hbar c/z_{0}$ & $z_{T} \ll z_{0}$ & $T_{z} \ll T$ \\ 
 \hline
\end{tabular}
\caption{Three pairs of limits and their associated conditions \cite{buhmann_ii}.}
\label{table:Limiting cases}
\end{table}
There are three pairs of characteristic quantities, of which three are related to distances and the others to temperatures. The first pair is the geometric distance $z_{0}$ and the geometric temperature, $T_{z} = \hbar c/(z_{0}k_{B})$, which is the temperature of radiation whose wavelength is of the order $z_{0}$. The second pair of parameters is the spectroscopic length, $z_{\omega} = c/\omega_{\pm}$, which is the measure of the wavelength of the maximum or minimum of the relevant atomic transition frequencies, $\omega_{\pm}$, and the spectroscopic temperature $T_{\omega} = \hbar\omega_{\pm}/k_{B}$. The last pair is the thermal length $z_{T} = \hbar c /(k_{B}T)$ and the environment temperature $T$. Comparing the geometric quantities with the spectroscopic quantities allows us to determine the retarded and non-retarded limits, and comparing the spectroscopic quantities with the thermal quantities allows us to determine the spectroscopic high- and low-temperature limits. Finally, comparing the geometric quantities with the thermal quantities allows us to determine the geometric high- and low-temperature limits as listed in Table \ref{table:Limiting cases}. It is impossible to simultaneously realize the retarded, spectroscopic high-temperature and geometric low-temperature limit; the same is true for the non-retarded, spectroscopic low-temperature, and geometric high-temperature limit.

\subsection{Scaling Relations}
\label{sec:Scaling Relations}
Since the main focus of this work is on Rydberg atoms, we will evaluate the scaling of the CP potential with respect to the principal quantum number $n$ (without showing graphical results). As shown in Table \ref{table:Scaling law}, the non-resonant term (as shown in Eq. \eqref{eq:U_CP^dip_nres}) scales with $n^{7}$ since the polarizability scales with $n^{7}$. Note that the Green's tensor as a function of Matsubara frequencies does not scale with the principal quantum number. The resonant term has four quantities that scale with $n$ and their combined contribution leads to an overall $n^{7}$, similar to the non-resonant term. Surprisingly, the total CP potential, $U^{\mathrm{nres}}_{u} + U^{\mathrm{res}}_{u}$, follows an $n^{4}$ power law due to their opposite signs.
\begin{table}[H]
\begin{tabular}{ |c|c| } 
 \hline
\textbf{Quantity} & \textbf{Power} \\ 
 \hline
 Polarizability $\boldsymbol{\alpha}_{u}$ & $n^{7}$ \\ 
 \hline
 Atomic transition frequencies $\omega_{uk}$ & $n^{-3}$ \\ 
 \hline
  Thermal photon number $N(\omega_{uk})$ & $n^{3}$ \\ 
 \hline
  Dipole moment $\textbf{d}_{uk}$ & $n^{2}$ \\ 
 \hline
  Green's tensor $\mathbf{G}^{(\mathrm{s})}(\omega_{uk})$ & $n^{6}$ \\ 
 \hline
  CP potential due to dipole interactions $U_{\mathrm{CP}}$ & $n^{4}$ \\ 
 \hline
\end{tabular}
\caption{Scaling of the CP potential and its associated quantities.}
\label{table:Scaling law}
\end{table}

\subsection{CP variations with atom-surface distances}
\begin{figure}[ht]
\includegraphics[width = 0.8\linewidth]{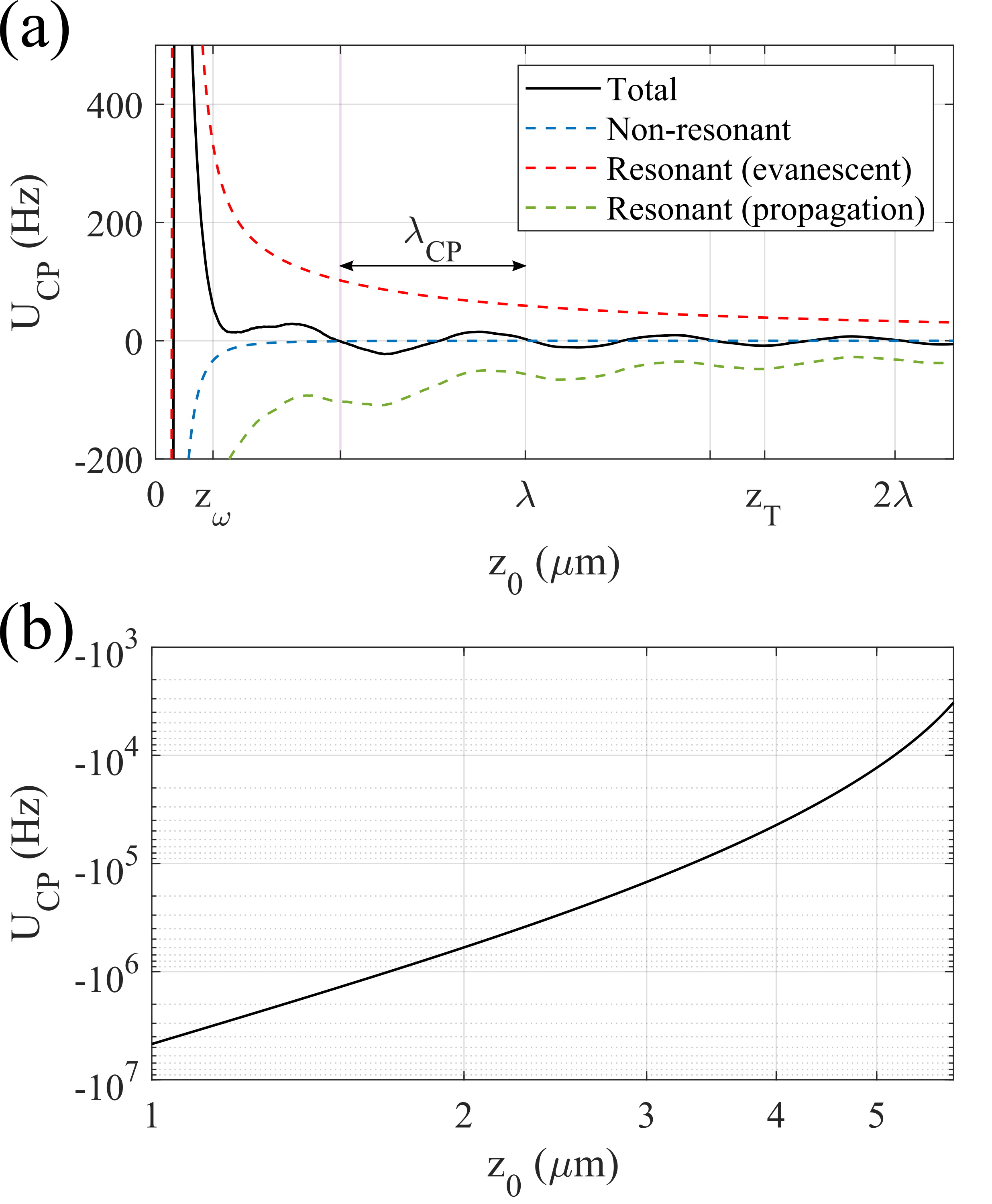}
\caption{(a) Linear scale plots of the total CP potential (solid-black curve), the non-resonant contribution (dashed-blue curve), the resonant contribution due to evanescent waves (dashed-red curve) and the resonant contribution due to propagating waves (dashed-green curve) of \textsuperscript{87}Rb in the 15S state versus $z_{0}$; (b) Log-log scale plots of the total CP potential calculated versus atom-surface separation. Note that $\lambda = \SI{139}{\mu}$m is roughly the wavelength of the 15S-14P transition, which is the dominant transition in this case. The parameters are: $T = \SI{10}{K}$, $z_{\omega} = \SI{21.6}{\mu}$m and $z_{T} = \SI{229}{\mu}$m.}
\label{fig:15S_Gr_VariationwithDist}
\end{figure}

Since the CP potential arises from the interaction between atoms and electromagnetic waves, we can expect that the behaviour of the CP potential in space can be split into the retarded and non-retarded regimes. Mathematically, the spatial variation of the CP potential is determined by the Green's function. At small atom-surface distances, the interactions with the evanescent waves will dominate and, at large atom-surface distances, the interactions with propagating waves will dominate. The Green's function scales with a $1/z_{0}^{3}$ power law in the non-retarded regime, and hence the same scaling is found in the CP potential.\\

As an example, Fig. \ref{fig:15S_Gr_VariationwithDist} shows the variation of the CP potential of the 15S state calculated versus $z_{0}$ and plotted on (a) a linear scale and (b) a log-log scale. Note that the wavelength of the 15S-14P transition, which is the dominant transition in this case, is $\lambda \approx \SI{139}{\mu}$m, and that the two characteristic lengths are $z_{\omega} \approx \SI{22}{\mu}$m and $z_{T} \approx \SI{229}{\mu}$m. The spatial variation of the non-resonant contribution is monotonic, similar to that of ground-state atoms. The figure reveals the dominance of the non-resonant contribution together with the resonant contribution due to evanescent waves at the distances much smaller than $z_{\omega}$ (below $\SI{6}{\mu}$m). For higher atom-surface separations, up to about $z_{\omega}$, the positive contribution from the resonant term (evanescent wave) dominates. From $z_{0} \sim \lambda/2$ onward, the resonant term due to the propagating-wave interactions comes into play and the CP potential oscillates around zero. The log-log plot in panel (b) also shows a clear distinction of the non-retarded regime, in which $U_{\mathrm{CP}} \propto 1/z_{0}^{3}$.\\

\begin{figure}[ht]
\includegraphics[width = \linewidth]{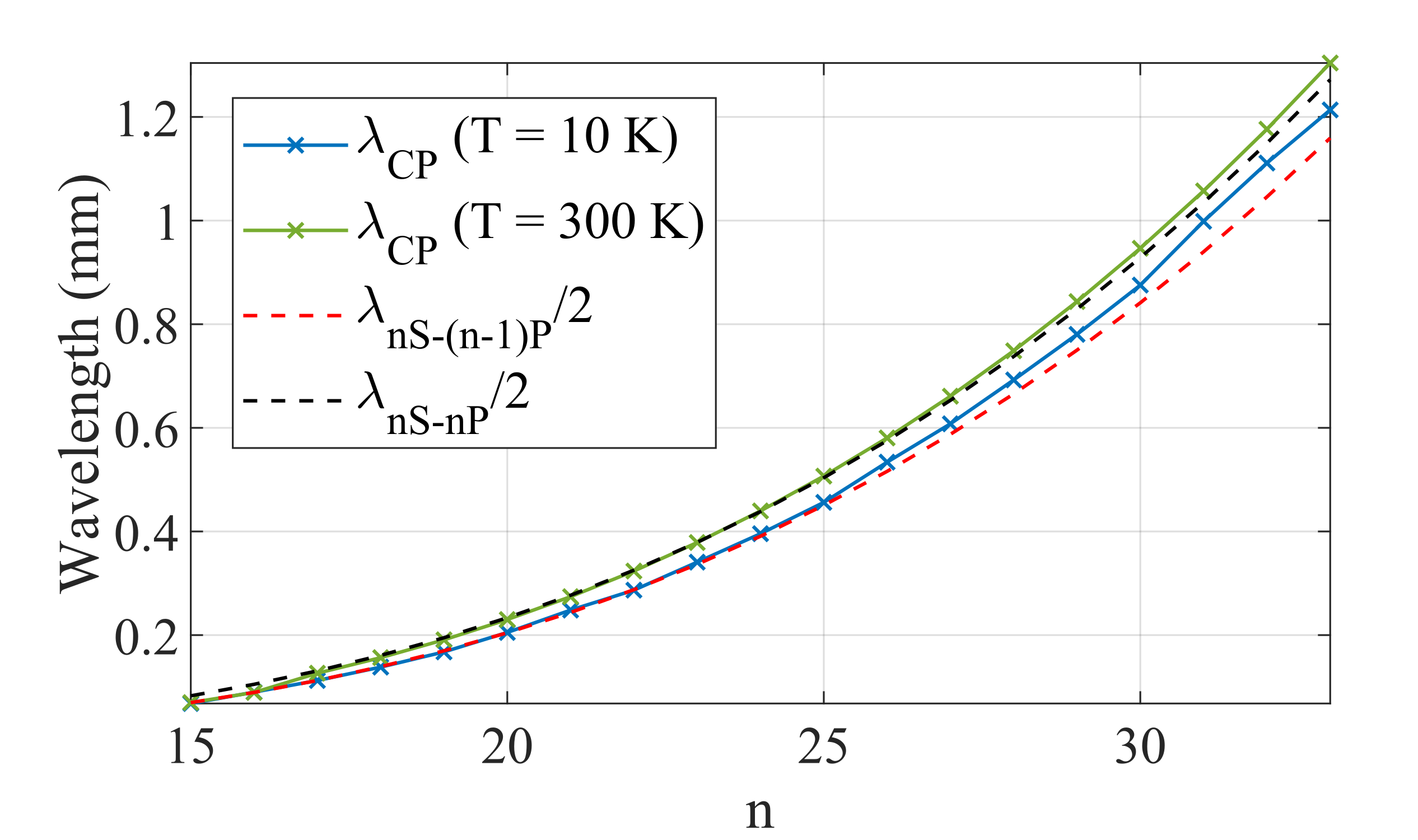}
\caption{Characteristic wavelengths of the CP potential calculated versus $n$ for $T = \SI{10}{K}$ (blue curve) and $T = \SI{300}{K}$ (green curve) calculated from the first cycle that appears after the atom-surface distance exceeds $\lambda_{n\mathrm{S}-(n-1)\mathrm{P}}/2$. Also shown are the half wavelengths of the nearest downward (red dashed curve) and upward (black dashed curved) atomic transitions calculated versus principal quantum numbers. At $10$ K, the wavelengths of the oscillation of the CP potential for $n$ roughly below $30$ leans towards the half wavelengths of the downward atomic transitions to the $(n-1)\mathrm{P}$ states, but they shift towards the half wavelengths of the upward atomic transitions to the $n\mathrm{P}$ states for most the states considered here due to increases in thermal photons (see discussion in Sec. \ref{Sec:Variations with temperature and principal quantum number}).}
\label{fig:lambdaCP}
\end{figure}

\begin{figure}[ht]
\includegraphics[width = \linewidth]{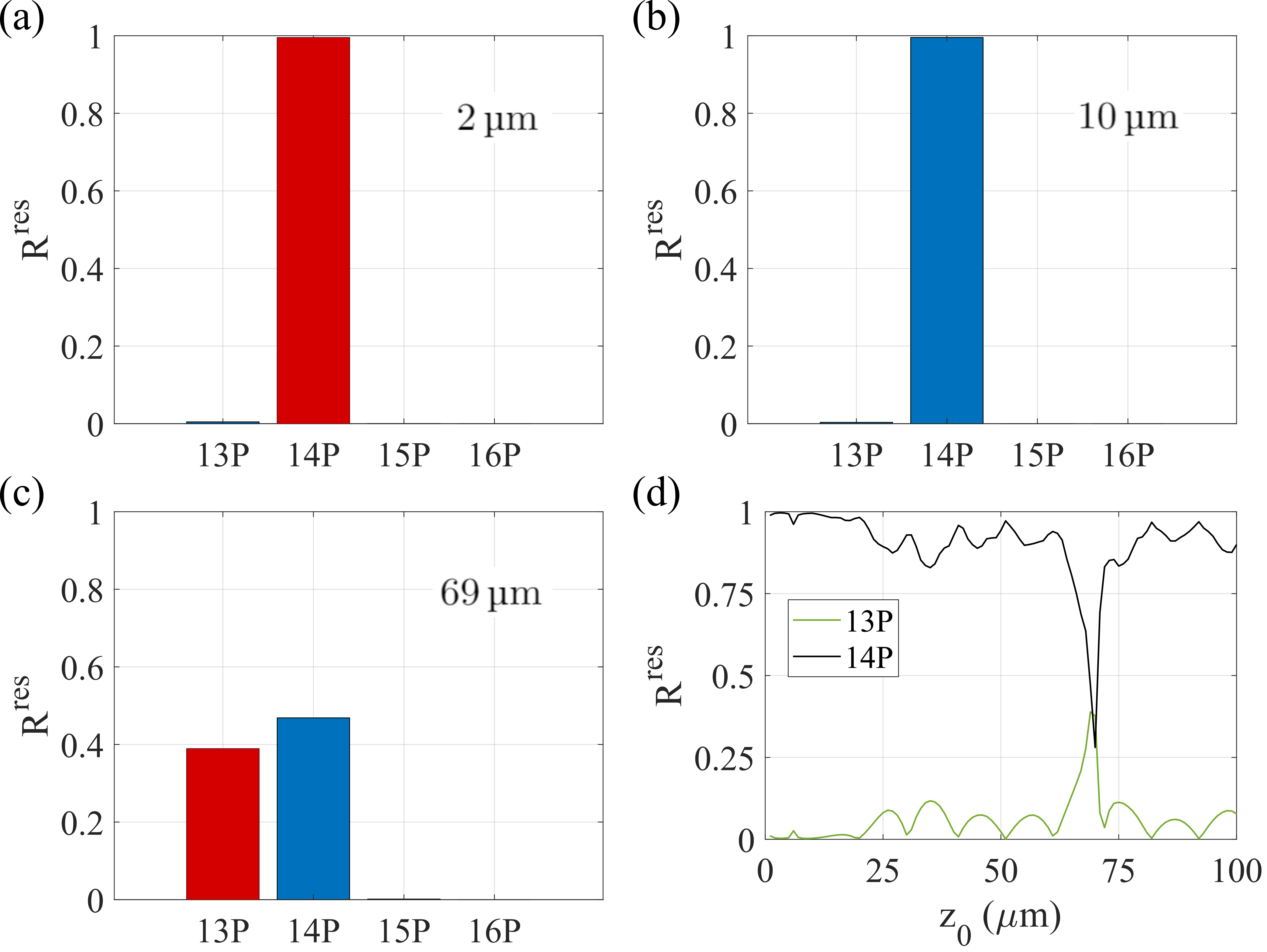}
\caption{Relative contributions of the four adjacent energy-states to the resonant term for an \textsuperscript{87}Rb atom in the 15S state ($z_{\omega} \approx \SI{22}{\mu}$m) at $z_{0} = $ (a) $\SI{2}{\mu}$m, (b) $\SI{10}{\mu}$m and (c) $\SI{69}{\mu}$m; (d) $R^{\mathrm{res}}$ versus atom-surface distances. Red and blue bars indicate negative and positive contributions, respectively. We can see that the signs of $R^{\mathrm{res}}$ for each transition change with distance, and at the points governed by $\lambda_{\mathrm{CP}}$ the positive and negative contributions cancel out each other, resulting in an overall zero potential as shown in Fig. \ref{fig:15S_Gr_VariationwithDist}. Note that, at this temperature, the contribution from the 15S-15P upward-transition is very small due to the lack of thermal photons. The parameters are $T = \SI{10}{K}$ and $z_{T} = \SI{229}{\mu}$m.}
\label{fig:15S_Gr_relative_con}
\end{figure}

In the retarded regime, the CP potential exhibits oscillatory behavior determined by the dominant atomic transition frequencies: the wavelength of the spatial oscillation of the CP potential, $\lambda_{\mathrm{CP}}$, is roughly half the wavelength of the dominant transition frequencies, $\lambda_{n\mathrm{S}-n\mathrm{P}}$ and $\lambda_{n\mathrm{S}-(n-1)\mathrm{P}}$ (see Fig. \ref{fig:lambdaCP}). The oscillation starts when the atom-surface distance is approximately half the corresponding atomic transition wavelength, which can be seen, for example, at the distance indicated by the purple vertical line in Fig. \ref{fig:15S_Gr_VariationwithDist}. For low-$n$ states, the dominant transition frequencies are high, the CP potential starts to oscillate at short atom-surface distances around $\SI{100}{\mu}$m; for high-$n$ states, the dominant transition frequencies are low with corresponding wavelengths of a few mm.\\ 
\\
In order to quantify how each intermediate state $\ket{k}$ in Eq. \eqref{eq:U_CP^dip_res} contributes to the total resonant CP potential, we consider a relative contribution, $R^{\mathrm{res}}$, which is defined as the absolute value of each term (specified by $k$) divided by the summation of the absolute value of all terms in Eq. \eqref{eq:U_CP^dip_res}. The relative contributions of the four adjacent energy-states for \textsuperscript{87}Rb atom in the 15S state ($z_{\omega} \approx \SI{22}{\mu}$m) at $z_{0} = \SI{2}{\micro\metre}, \SI{10}{\mu}$m, and $\SI{69}{\mu}$m together with $R^{\mathrm{res}}$ for the 13P and 14P states are shown versus atom-surface distances in Fig. \ref{fig:15S_Gr_relative_con}. Since we know that the non-resonant potential is monotonic, the underlying mechanics of the oscillations of the CP potential must originate from the resonant term. The signs of the individual contributions alter with distances and at some points the positive and negative contributions cancel out each other, resulting in a negligible potential. We can also see that $z_{\omega} \ll z_{T}$, indicating that this is the spectroscopic low-temperature regime. As a result, the contributions from all upward atomic transitions are negligible due to the lack of thermal photons. 

\subsection{Temperature and Rydberg state dependence}
\label{Sec:Variations with temperature and principal quantum number}
Thermal effects on the CP potential of ground-state atoms and molecules depend on the dominant atomic transition frequencies, which determine the non-retarded and retarded types: the near-surface (order of $\mu$m) CP potentials of molecules are likely to be non-retarded and hence insensitive to temperature as the dominant transition frequencies are low ($\sim \SI{1e13}{rad/s}$), while the CP potentials for ground-state atoms are strongly retarded and thus greatly affected by temperature since the dominant transition frequencies are high ($\sim \SI{1e15}{rad/s}$) \cite{Ellingsen_Scheel_2009, Ellingsen_Scheel_2010, Klim_Most_2018}.\\
\\
First, we consider how the conductivity of graphene depends on temperature. There is an explicit linear temperature dependence in the non-resonant potential--see Eq. \eqref{eq:U_CP^dip_nres}. The temperature dependence for the resonant term, Eq. \eqref{eq:U_CP^dip_res}, is embedded in the thermal photon distribution function, where $N(\omega_{uk}) \approx k_{B}T/\hbar\omega_{uk}$ as long as $\hbar\omega_{uk}/k_{B}T \ll 1$, i.e. the spectroscopic high-temperature limit. On the other hand, there are five quantities that depend on the principal quantum number as mentioned in Section \ref{sec:Scaling Relations}.\\

\begin{figure}[ht]
\includegraphics[width = \linewidth]{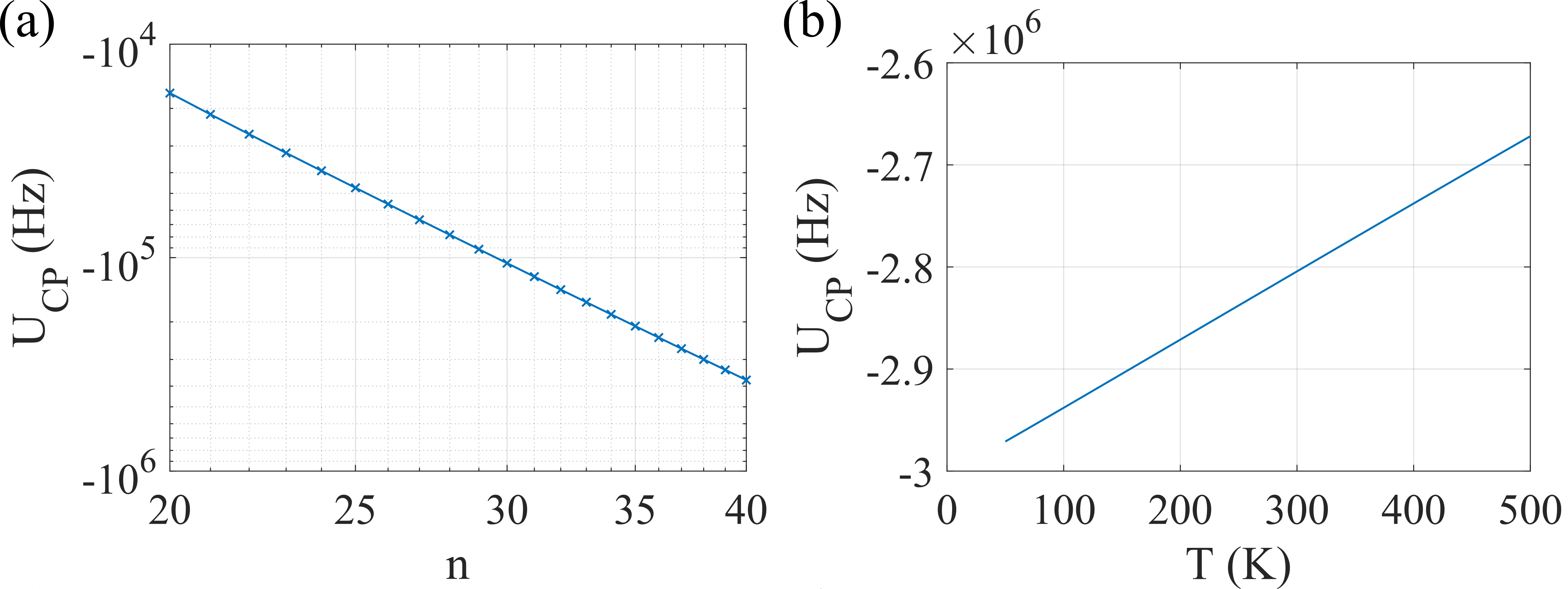}
\caption{Variations of the CP potential with (a) principal quantum number, calculated at fixed distance $z_{0} = \SI{10}{\mu}$m and $T = \SI{10}{K}$ with $T_{z} = \SI{229}{K}$ and (b) temperature, calculated for the 40S state at $z_{0} = \SI{5}{\mu}$m ($T_{z} = \SI{458}{K}$ and $T_{\omega} \approx \SI{3}{K}$). The CP potential is proportional to $n^{4}$ and $T$.}
\label{fig:Variations_with_n_T}
\end{figure}
Fig. \ref{fig:Variations_with_n_T} shows the variations of the CP potential with (a) principal quantum number, calculated at $z_{0} = \SI{10}{\mu}$m and (b) temperature, calculated for the 40S state at $z_{0} = \SI{5}{\mu}$m ($T_{z} = \SI{458}{K}$ and $T_{\omega} \approx \SI{3}{K}$).
Even though, the non-resonant and resonant potential, each scales with $n^{7}$, the total CP potential obeys an $n^{4}$ power law. The CP potential changes linearly with temperature. The reason is that the dominant transition energies of Rydberg atoms between two nearest energy-states, $(n-1)$P and $n$P, are small compared to thermal energy, $k_{B}T$, which makes the condition $\hbar\omega_{uk}/k_{B}T \ll 1$ so the average photon number is reduced to $N(\omega_{uk}) \approx k_{B}T/\hbar\omega_{uk}$, confirming a linear relation with $T$ as shown in panel (b). \\

\begin{figure}[ht]
\includegraphics[width = \linewidth]{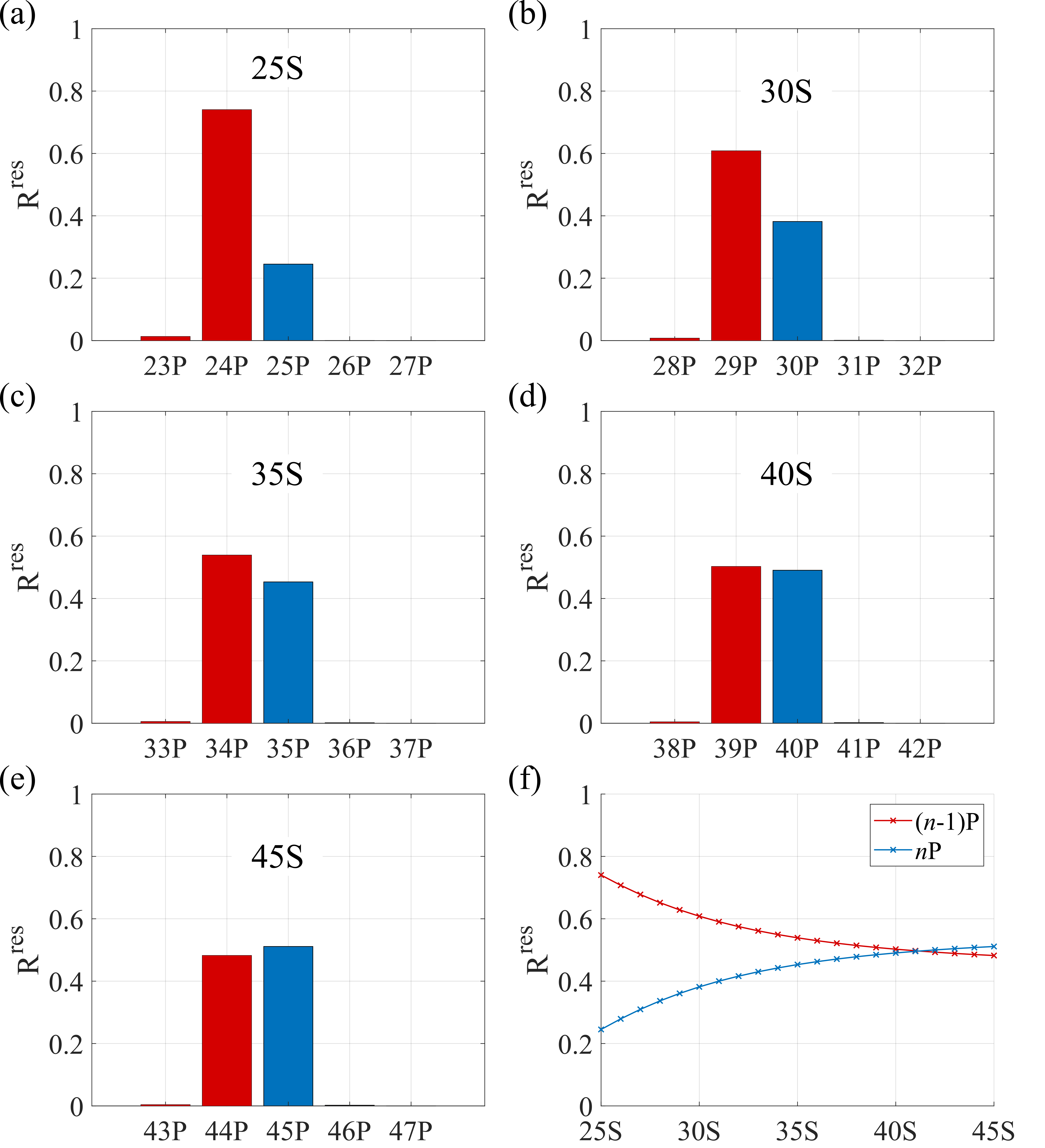}
\caption{The relative contributions of the intermediate states to the resonant parts of the CP potential at $\SI{5}{\mu}$m for an $^{87}$Rb atom in (from (a)-(e) respectively) the 25S, 30S, 35S, 40S and 45S states. Red and blue bars indicate negative and positive contributions, respectively. In this case, $T_{\omega} \approx \SI{14}{K}$ for the 25S state and $\approx \SI{2}{K}$ for the 45S state, and $T_{z} = \SI{458}{K}$, indicating the non-retarded limit. Comparing the bar heights among the $(n-1)$P states and among the $n$P states, we can see that the contributions from states with energy levels below the target states get smaller as $n$ increases and vice versa for the higher energy states. The temperature is $T = \SI{10}{K}$.}
\label{fig:Res_contributions_5_nS}
\end{figure}

We now consider how the resonant contributions change with the principal quantum number. In Fig. \ref{fig:Res_contributions_5_nS}, we plot the contributions of the intermediate states to the resonant parts of the CP potential in the 25S, 30S, 35S, 40S and 45S states for the environment temperature $T = \SI{10}{K}$. Note that the systems are in the non-retarded limit, and the spectroscopic temperature for the 25S state is approximately $\SI{14}{K}$, slightly higher than the environment temperature, whereas $T_{\omega} \approx \SI{2}{K}$ for the 45S state, indicating the spectroscopic high-temperature limit. We can see that the contributions from the intermediate states $\ket{k}$ to the resonant parts also change when the principal quantum number of the target state $\ket{u}$ changes.
As shown in Fig. \ref{fig:Res_contributions_5_nS}, the (downward-transition) contributions from the intermediate states with energy levels below the target states (below and including $(n-1)$P state) get smaller as $n$ increases and vice versa for the intermediate states with higher energies (i.e. upward-transition contributions), which is captured in panel (f). In other words, the system approaches the spectroscopic high-temperature limit as $n$ increases, resulting in an increase in the repulsive resonant potential. This behaviour of the resonant parts will affect the scaling of the CP potential with $n$ and also enhances the temperature dependence of the upward-transition contributions as we shall see in more detail in Sec. \ref{sec:Fitted empirical functions}.\\

\begin{figure}[ht]
\includegraphics[width = \linewidth]{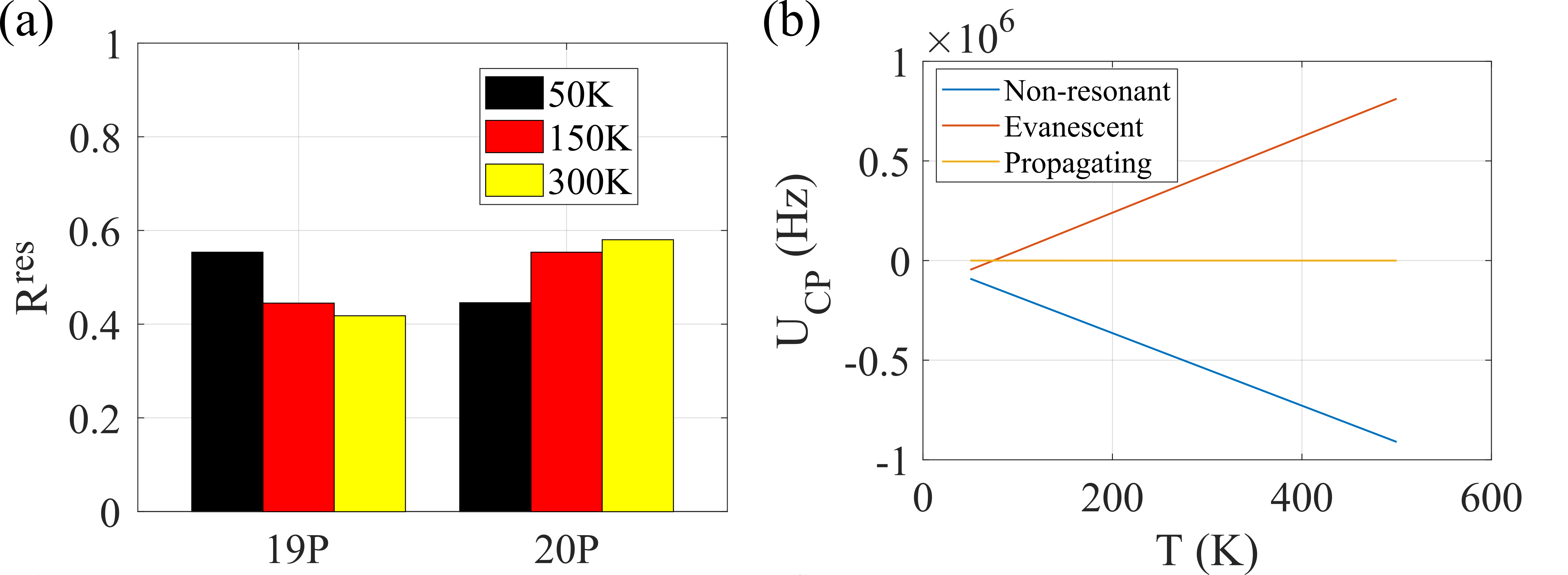}
\caption{(a) Relative contributions from the 19P and 20P states; (b) the non-resonant, evanescent-wave, and propagating-wave components of the CP potential for an atom in the 20S state at $\SI{5}{\mu}$m ($T_{\omega} = \SI{35}{K}$). Panel (a) shows that the positive contribution from the 20P state increases, while the negative contribution from the 19P state decreases as $T$ increases, resulting in a more positive evanescent-wave potential as $T$ increases, as shown in panel (b).}
\label{fig:20S_Gr_Bar_vsT}
\end{figure}

We now proceed to look into detail at how the resonant contributions from the two nearest energy states change with temperature. For the 20S state, there is a slight increase in the positive contribution from the 20P state as $T$ increases, while the opposite happens to the 19P state, as shown in Fig. \ref{fig:20S_Gr_Bar_vsT}(a), resulting in the evanescent-wave potential being more positive as $T$ increases. Fig. \ref{fig:20S_Gr_Bar_vsT}(b) shows the non-resonant and resonant contributions due to evanescent and propagating waves. The attractive contribution from the non-resonant part is stronger than the repulsive resonant potential.

\subsection{Effects of changing Fermi energy}
\begin{figure}[ht]
\includegraphics[width = \linewidth]{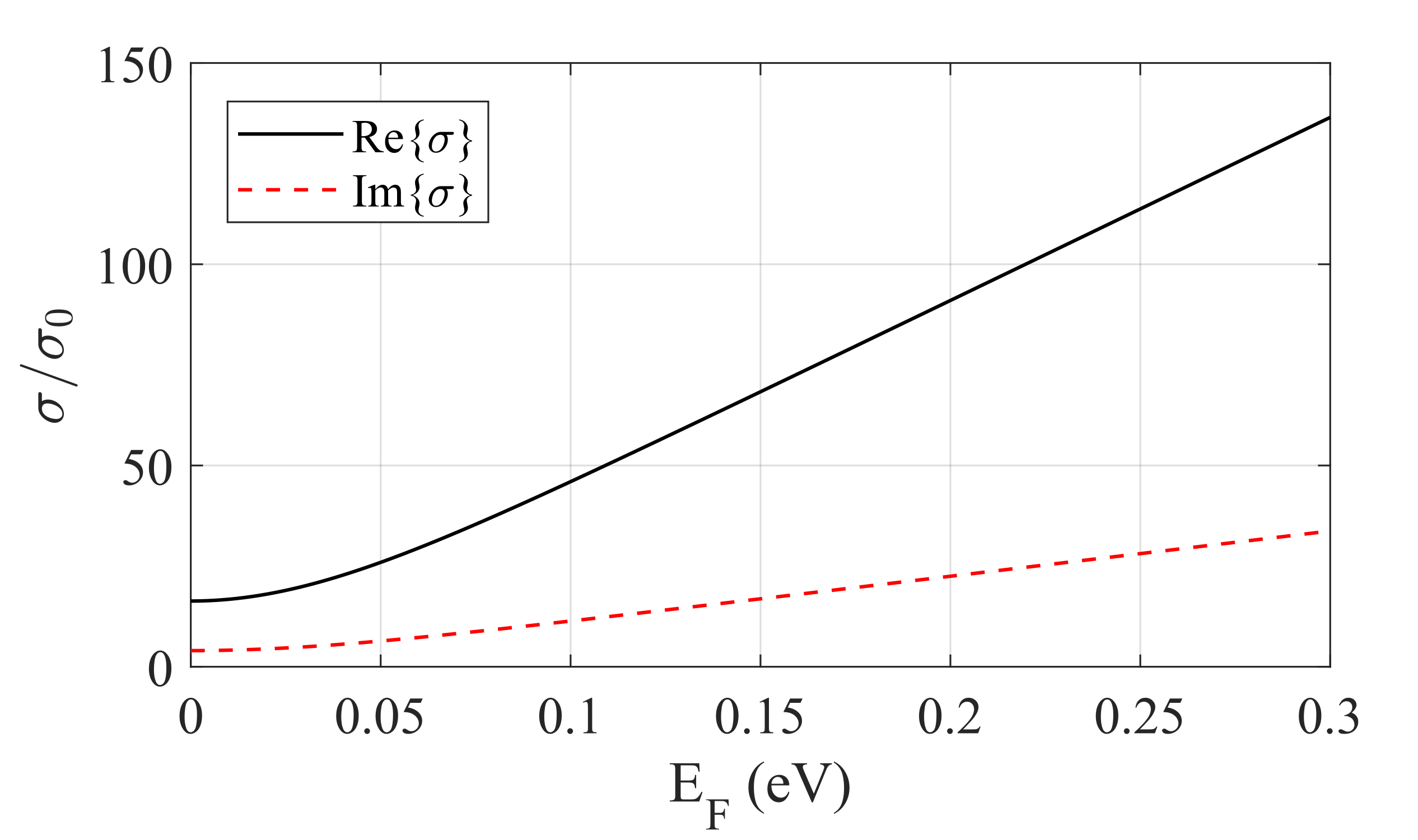}
\caption{The real and imaginary parts of the optical conductivity of graphene, calculated at $\omega = \SI{9.88e11}{rad\per\second}$, which is the atomic transition frequency between the 30S and 30P state. The conductivity increases almost linearly as the Fermi energy increases. The temperature is $T = \SI{300}{K}$.}
\label{fig:sigma_vs_E_F}

\end{figure}

\begin{figure}[ht]
\includegraphics[width = \linewidth]{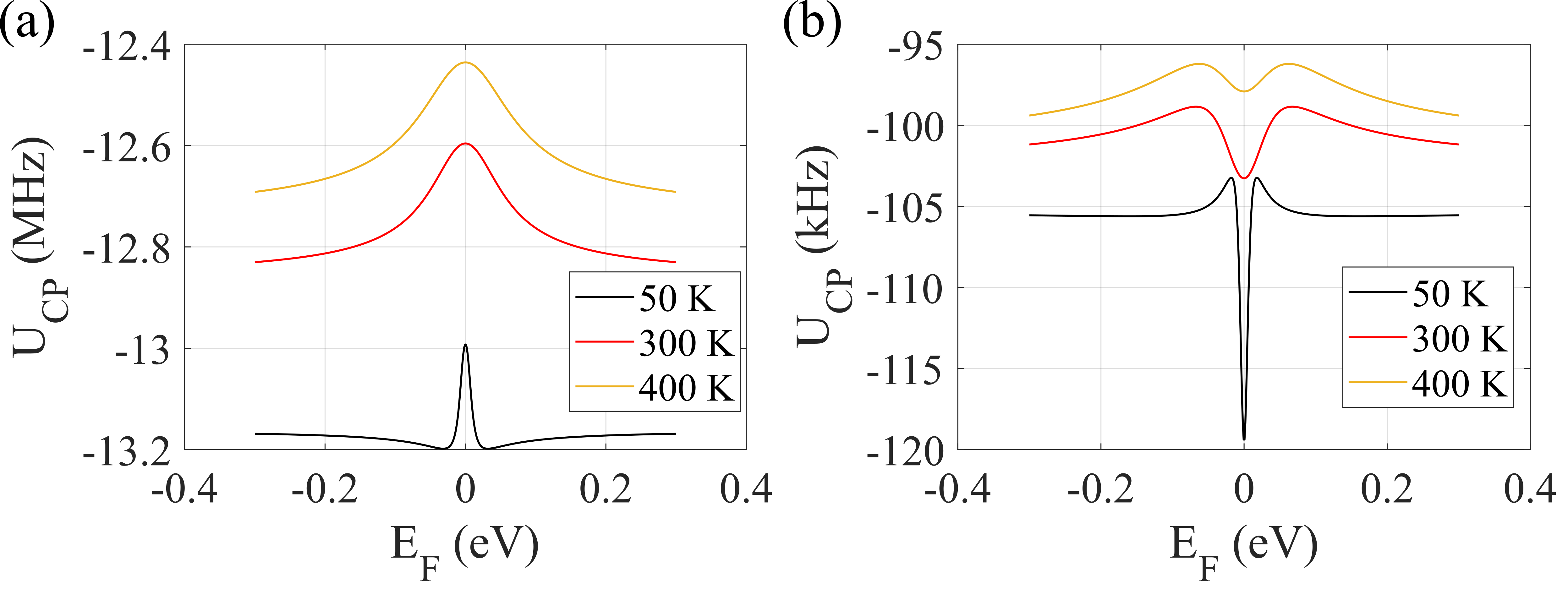}
\caption{CP potential calculated versus Fermi energy, $E_{F}$, at atom-surface distances (a) $z_{0} = \SI{2}{\mu}$m and (b) $z_{0} = \SI{10}{\mu}$m for three different temperatures for a \textsuperscript{87}Rb atom in the 30S state.}
\label{fig:30S_vs_E_F}
\end{figure}

In this section, we will investigate how the CP potential of single-layer graphene depends on its Fermi energy, which influences the conductivity of the graphene sheet and, hence, its reflection coefficients and underlying Green's tensor. We expect stronger non-resonant contributions to the CP potential as we increase the Fermi energy since we are adding more conduction electrons. Regarding the resonant part, the dominant atomic transition frequencies of the atomic states usually considered are in the microwave-terahertz spectral regions, which are small compared to the typical Fermi energy. Consequently, as discussed in Appendix \ref{appendix:Graphene's Optical Properties}, we expect that graphene's optical conductivity will be dominated by the intraband-process term, which is almost linear in $E_F$ and is a few orders of magnitude higher than the universal AC conductivity of graphene $\sigma_{0}$, as shown in Fig. \ref{fig:sigma_vs_E_F}. However, the CP potential does not vary linearly with the Fermi energy as we can see in Fig. \ref{fig:30S_vs_E_F}, where we show the CP potential calculated at (a) $z_{0} = \SI{2}{\mu}$m and (b) $z_{0} = \SI{10}{\mu}$m in the 30S state. In (a), at $T = \SI{50}{K}$, the potential peaks at $E_{F} = \SI{0}{eV}$ then decreases and rises again as the Fermi energy deviates from $0$ eV. This is not the case for $T = \SI{300}{K}$ and $T = \SI{400}{K}$: the potential then gets more attractive as the doping level increases. In (b), all three curves exhibit similar behaviors: the potential is the most attractive at $E_{F} = \SI{0}{eV}$.

\section{Fitted empirical functions}
\label{sec:Fitted empirical functions}
In this section, we will try to find fitted empirical functions that follow the power laws discussed in the preceding sections.
Let us begin by writing the CP potential in the form:
\begin{equation}
    U_{\mathrm{CP}} = -\frac{C_{\alpha}}{z_{0}^{\alpha}},
    \label{eq:U_scaling_law}
\end{equation}
where $C_{\alpha}$ is a positive dispersion coefficient and $\alpha$ is a scaling power to be evaluated. 
\subsection{Low-temperature limit}
\label{sec:Low-temperature limit}
\begin{figure}[ht]
\includegraphics[width = 0.8\linewidth]{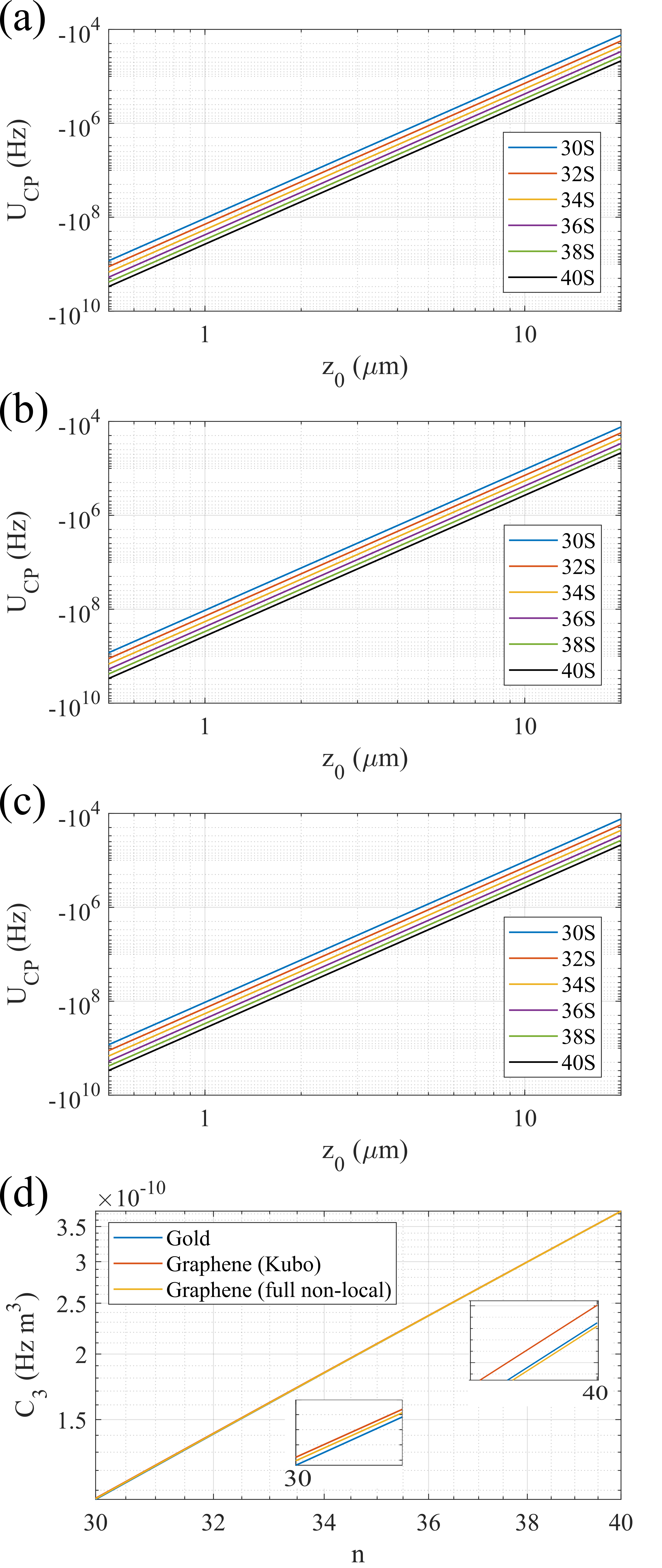}
\caption{Panels (a)-(c) show the CP potential calculated, versus atom-surface separation, and plotted on a log-log scale for a $1$-$\mu$m-thick gold sheet, graphene modeled by the Kubo conductivity and graphene modeled by the full non-local conductivity, respectively. Panel (d) shows the relations between the dispersion coefficients, $C_{3}$ (obtained from Eq. \eqref{eq:U_scaling_law}), and the principal quantum numbers of the atomic states considered, calculated at $z_{0} = \SI{10}{\mu}$m, with two insets zoomed near $n = 30$ and $n = 40$.}
\label{fig:Fitted_function_2}
\end{figure}

We first consider the non-retarded, spectroscopic low-temperature limit in order to determine the scaling relation of the CP potential that largely depends on the atom-surface separation and on the principal quantum number. Figure \ref{fig:Fitted_function_2} shows the CP potential at $T = \SI{10}{K}$ for (a) a $\SI{1}{\mu}$m-thick gold sheet, (b) graphene modeled by the Kubo conductivity and (c) graphene modeled by the full non-local conductivity. Panel (d) shows the dispersion coefficients, $C_{3}$, versus the principal quantum numbers. From the slopes of the CP potential in panels (a)-(c), $\alpha$ is determined to be $3$ and it follows from the slopes of $C_{3}$ in panel (d) that $C_{3} \propto n^{4}$ approximately. To be more precise, assuming $C_{3} = q_{1}n^{4} + q_{2}n^{3}$ ($q_{1}$ and $q_{2}$ are coefficients), we find, for the gold sheet, 
\begin{equation}
    \frac{C_{3}}{(\mathrm{Hz\cdot m^3})} = (\SI{1.936e-16}{})n^{4} - (\SI{1.893e-15}{})n^{3},
    \label{eq:C3_gold}
\end{equation}
for graphene modeled by the Kubo conductivity,
\begin{equation}
    \frac{C_{3}}{(\mathrm{Hz\cdot m^3})} = (\SI{1.923e-16}{})n^{4} - 
    (\SI{1.840e-15}{})n^{3},
    \label{eq:C3_Kubo}
\end{equation}
and for graphene modeled by the full non-local model,
\begin{equation}
    \frac{C_{3}}{(\mathrm{Hz\cdot m^3})} = (\SI{1.924e-16}{})n^{4} - (\SI{1.848e-15}{})n^{3}.
    \label{eq:C3_fnl}
\end{equation}\\

Alternatively, if we assume that $C_{3}$ is proportional to a single power of $n$, we find, for the gold sheet, 
\begin{equation}
    \frac{C_{3}}{(\mathrm{Hz\cdot m^3})} = (\SI{3.394e-17}{})n^{4.397},
    \label{eq:C3_gold_2}
\end{equation}
for graphene modeled by the Kubo conductivity,
\begin{equation}
    \frac{C_{3}}{(\mathrm{Hz\cdot m^3})} = (\SI{3.543e-17}{})n^{4.385},
    \label{eq:C3_Kubo_2}
\end{equation}
and for graphene modeled by the full non-local model,
\begin{equation}
    \frac{C_{3}}{(\mathrm{Hz\cdot m^3})} = (\SI{3.517e-17}{})n^{4.387}.
    \label{eq:C3_fnl_2}
\end{equation}
We can see from the above equations that both the Kubo and full non-local models give quantitatively similar results.

\subsection{Including temperature-dependent effects}

\begin{figure}[ht]
\includegraphics[width = \linewidth]{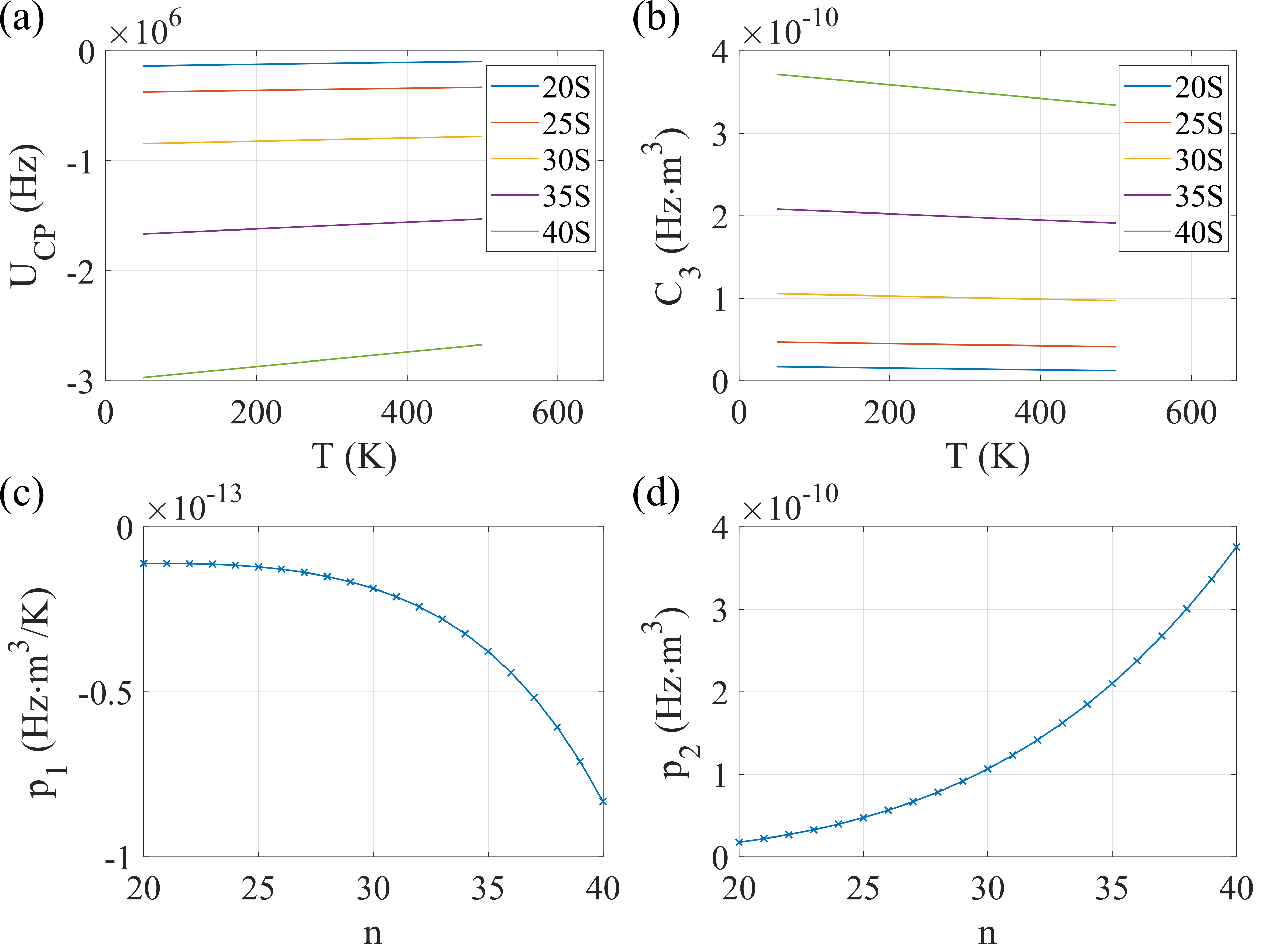}
\caption{(a, b) The CP potential and the dispersion coefficient calculated versus temperature for a graphene-monolayer modeled by the Kubo conductivity; (c, d) the polynomial coefficients of the fitted empirical formula. The spacing of $C_{3}$ follows an $n^{4}$ power law, while its slope follows an $n^{7}$ power law. The parameter is $z_{0} = \SI{5}{\mu}$m.}
\label{fig:Scaling_vs_T_n}
\end{figure}

In this subsection, we try to include the scaling law of the CP potential with temperature in an attempt to find a more complete fitted empirical formula which can describe the CP potential of a \textsuperscript{87}Rb atom near a single graphene sheet in the non-retarded regime and for various $n$. This is done by calculating the CP potential versus temperature to obtain the dispersion coefficient, $C_{3}$, as a function of temperature, $T$, and principal quantum number, $n$, in the form of a linear equation $C_{3}(n, T) = p_{1}(n)T + p_{2}(n)$. Figs. \ref{fig:Scaling_vs_T_n}(a) and (b) show the CP potential near graphene vs $T$ for the 20S, 25S, 30S, 35S and 40S states and its associated dispersion coefficient, $C_{3}$. The $U_{\mathrm{CP}}(T)$ curves show a linear relationship between the potential and temperature. However, the slopes of the $U_{\mathrm{CP}}(T)$ and $C_{3}(T)$ curves become steeper as $n$ increases since the positive resonant contributions increase as $n$ increases, as shown in Fig. \ref{fig:Res_contributions_5_nS}. The non-linear relations between the coefficients, $p_{1}$ and $p_{2}$, and $n$ are shown in Figs. \ref{fig:Scaling_vs_T_n}(c) and (d). We can fit them with polynomial equations of degree $7$ and $4$ as follows (see Sec. \ref{sec:Scaling Relations}):

% \begin{multline}
%     p_{1}(n) = \Big[-\SI{4.415e-19}{}n^{4} + \SI{3.91e-17}{}n^{3}\\
%     - \SI{1.376e-15}{}n^{2} + \SI{2.245e-14}{}n\\
%     + \SI{1.52e-13}{}\Big]\times (\mathrm{Hz\cdot m^3}/K),
%     \label{eq:p1 of C3}
% \end{multline}

% \begin{multline}
%     p_{2}(n) = \Big[\SI{1.982e-16}{}n^{4} - \SI{1.955e-15}{}n^{3}\\
%     - \SI{1.293e-14}{}n^{2} + \SI{3.565e-13}{}n\\
%     - \SI{2.819e-13}{}\Big]\times (\mathrm{Hz\cdot m^3}).
%     \label{eq:p2 of C3}
% \end{multline}\\

\begin{multline}
    \frac{p_{1}(n)}{(\mathrm{Hz\cdot m^3}/K)} = \Big[-\SI{4e-25}{}n^{7} - \SI{9.38e-15}{}\Big],
     \label{eq:p1 of C3_2}
\end{multline}

\begin{multline}
     \frac{p_{2}(n)}{(\mathrm{Hz\cdot m^3})} = \Big[\SI{1.866e-16}{}n^{4} - \SI{1.614e-15}{}n^{3}\Big] .
     \label{eq:p2 of C3_2}
\end{multline}\\
%1.866 1.614
We may write the fitted empirical function for the CP potential for graphene-monolayer in the spectroscopic high-temperature and non-retarded regime as

\begin{equation}
    U_{\mathrm{CP}}(n, T, z_{0}) = -\frac{p_{1}(n)T + p_{2}(n)}{z_{0}^{3}}.
    \label{eq:U_CP_empirical}
\end{equation}\\

\subsection{Accuracy of the fitted empirical functions}
\begin{figure}[ht]
\includegraphics[width = \linewidth]{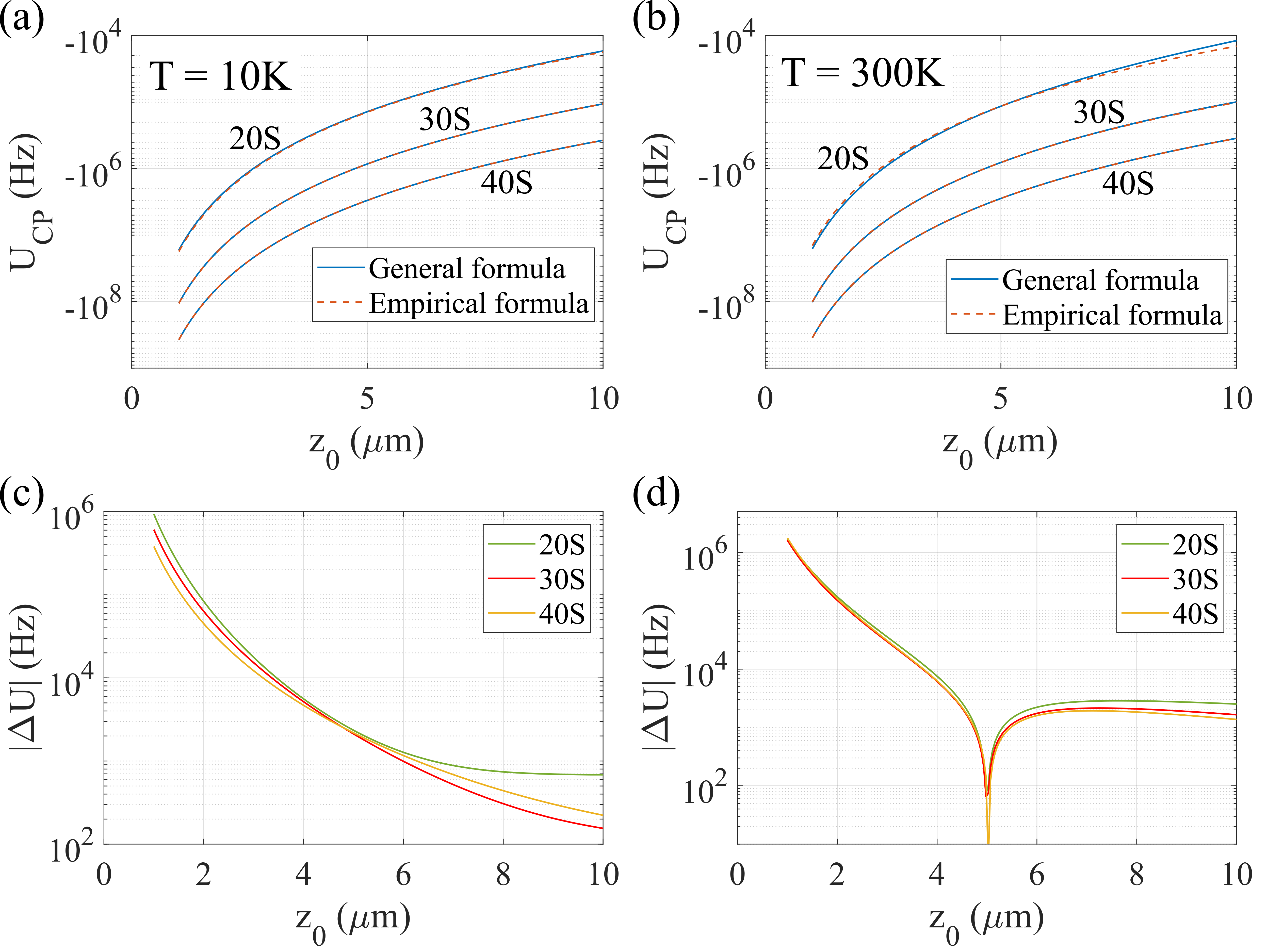}
\caption{The CP potential calculated by the general formula (blue solid curves) and by the fitted empirical function Eq.\eqref{eq:U_CP_empirical} (orange dashed curves) at (a) $T= \SI{10}{K}$ and (b) $T= \SI{300}{K}$ and their corresponding differences (c, d). The fitted function provides the least accurate results for the 20S state with a relative error as high as $10\%$. As for the other two states, the relative error is far below $1\%$. The parameter is $z_{0} = \SI{5}{\mu}$m.}
\label{fig:Accuracyoftheformula}
\end{figure}

We now check the accuracy of our fitted empirical formula, Eq.\eqref{eq:U_CP_empirical}, compared with the general formula, Eq. \eqref{eq:nres_res}, by plotting the CP potential for atom-surface distances in the range $\SI{1}{\mu}$m to $\SI{10}{\mu}$m and calculating the absolute values of the energy differences.
Fig. \ref{fig:Accuracyoftheformula} shows the CP potential calculated by the general formula (blue solid curves) and the empirical formula (orange dashed curves), together with their corresponding energy differences, $\abs{\Delta U}$, at (a, c) $T = \SI{10}{K}$ and (b, d) $T = \SI{300}{K}$. We can see that the empirical formula gives results that are off by $\sim \SI{1}{MHz}$ at $\SI{1}{\mu}$m. Beyond this point, the differences are of order smaller than Megahertz. The empirical formula provides the least accurate results for the 20S state with a relative error as high as $10\%$. As for the other two states, the relative error is far below $1\%$.\\

\section{Heterostructures containing two graphene layers}
\label{sec:Double-Layer Graphene Heterostructures}
\begin{figure}[ht]
\includegraphics[width = \linewidth]{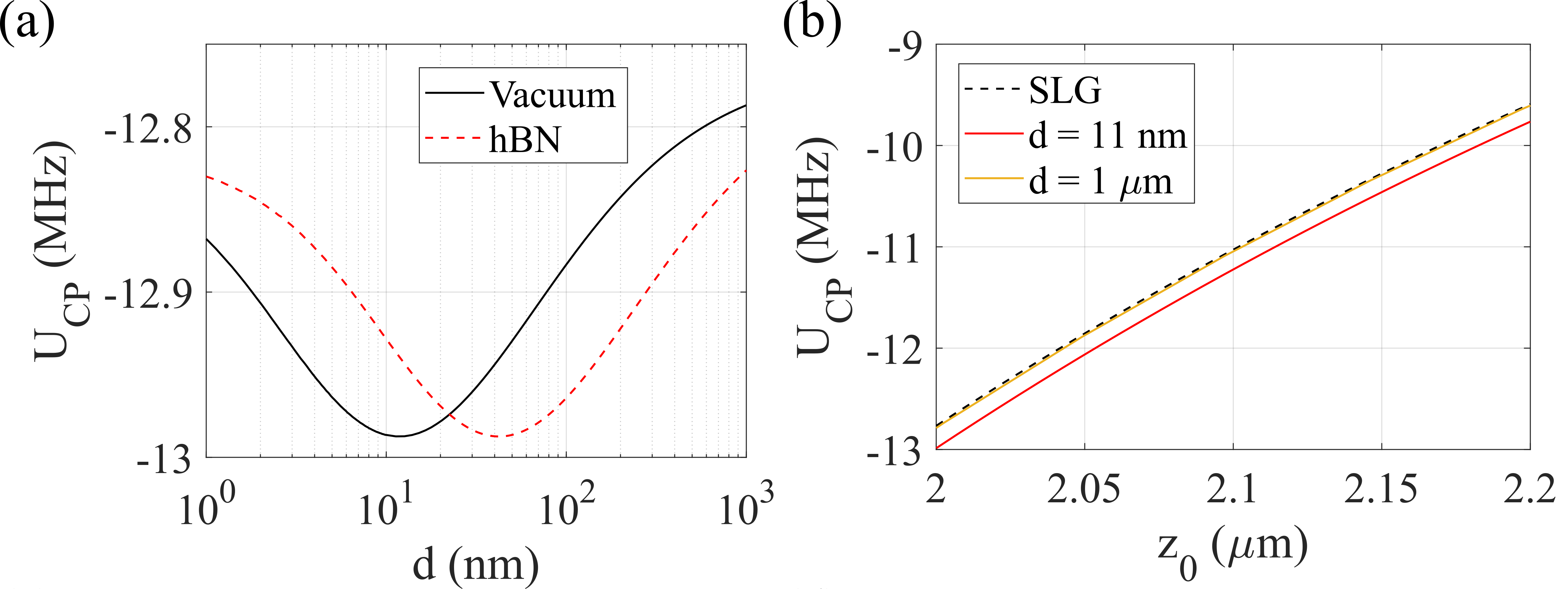}
\caption{CP potential of a \textsuperscript{87}Rb atom in the 30S state (a) versus the spacing $d$ of heterostructures containing two graphene layers, separated by vacuum and by hBN; (b) versus $z_{0}$ for a single-layer graphene (SLG, black dashed curve) and a graphene-vacuum-graphene structure with spacing $d = \SI{11}{nm}$ and $d = \SI{1}{\mu m}$. The parameters are: T = $\SI{300}{K}$, $E_{F} = \SI{0.1}{eV}$, $z_{0} = \SI{2}{\mu m}$ and  $\epsilon_{\mathrm{hBN}} = 3.58$.}
\label{fig:30S_spacing}
\end{figure}

\begin{figure}[ht]
\includegraphics[width = \linewidth]{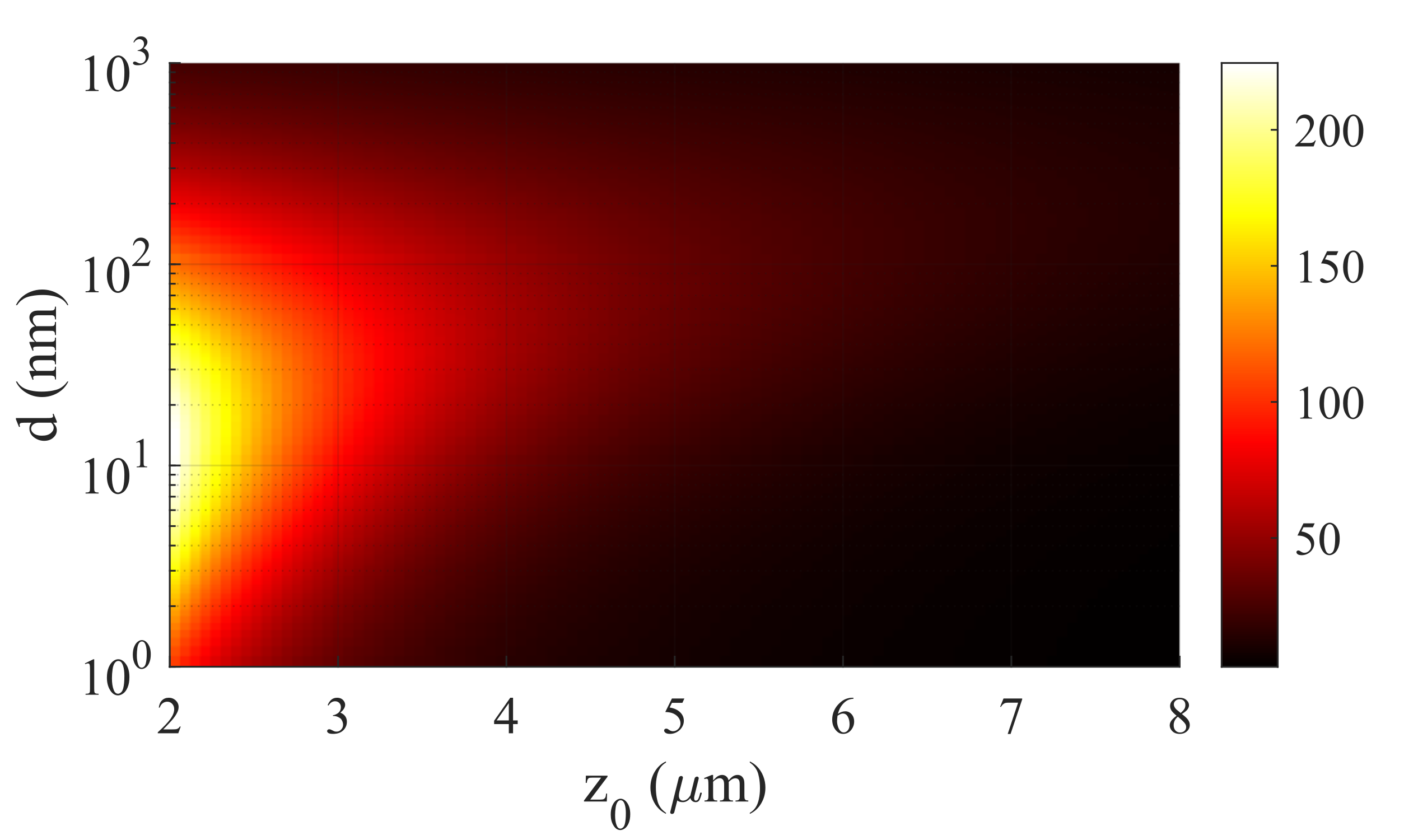}
\caption{Color map of the energy difference (in kHz) between the CP potential of single-layer graphene and graphene-vacuum-graphene structure with spacing $d$ calculated versus $z_0$ and $d$ (on a linear-log scale) for a \textsuperscript{87}Rb atom in the 30S state. The plot shows that the behavior of the two layers of graphene approaches that of SLG as the spacing $d$ increases. The parameters are: T = $\SI{300}{K}$, $E_{F} = \SI{0.1}{eV}$.}
\label{fig:30S_air_z_0_vs_d_T300_log}
\end{figure}

In this section, we study the CP potential of heterostructures containing two separated layers of graphene, which provide a wider range of tuneable optical properties than single-layer graphene (SLG) \cite{rodrigo_2017, woessner2014, brar_hybrid_2014}. 
Hexagonal boron nitride (hBN) is commonly used to integrate with graphene to form complex structures; we can use it, for example, as a substrate, a tunnel barrier \cite{geim_grigorieva_2013}, or an encapsulating layer \cite{thomsen_gunst_2017, Han_Pan_2019}. We can grow hBN on graphene vertically \cite{Yankowitz_2014} or laterally \cite{Wrigley_2021}.\\
\\
Let us consider a simple van der Waals (vdW) heterostructure comprising an hBN layer with permittivity $\epsilon_{\mathrm{hBN}} = 3.58$ \cite{laturia_dielectric_2018} sandwiched between two graphene monolayers with spacing $d$ between them. Bare graphene layers (i.e. without an hBN spacer) in vacuum will also be considered as a comparison. Figure \ref{fig:30S_spacing} shows the CP potential of the said structures at $T = \SI{300}{K}$ and $z_{0} = \SI{2}{\mu}$m for a \textsuperscript{87}Rb atom in the 30S state. In (a), the spacing $d$ is varied between $1$ nm and $\SI{1}{\mu}$m; the potential becomes more negative when we increase the spacing between two graphene layers until $d \approx \SI{11}{nm}$ for the graphene-vacuum-graphene structure and $d \approx \SI{73}{nm}$ for the graphene-hBN-graphene structure, when the potential gets less negative again. In (b), we show that the potential of the structure containing two layers of graphene approaches that of the single-layered one as $d$ is increased to $\SI{1}{\mu}$m.\\
\\
In order to show the relationship between $z_{0}$ and $d$, in Fig. \ref{fig:30S_air_z_0_vs_d_T300_log} we plot the color map of the differences (in kHz) between the CP potential of SLG and that of graphene-vacuum-graphene structure on a linear-log scale. We can see once again that the behavior of the two-graphene-layered structure approaches that of SLG as the spacing $d$ increases and that the atom effectively experiences the two-graphene-layered structure as SLG when the atom-surface distance is large enough.

\begin{figure}[ht]
\includegraphics[width = \linewidth]{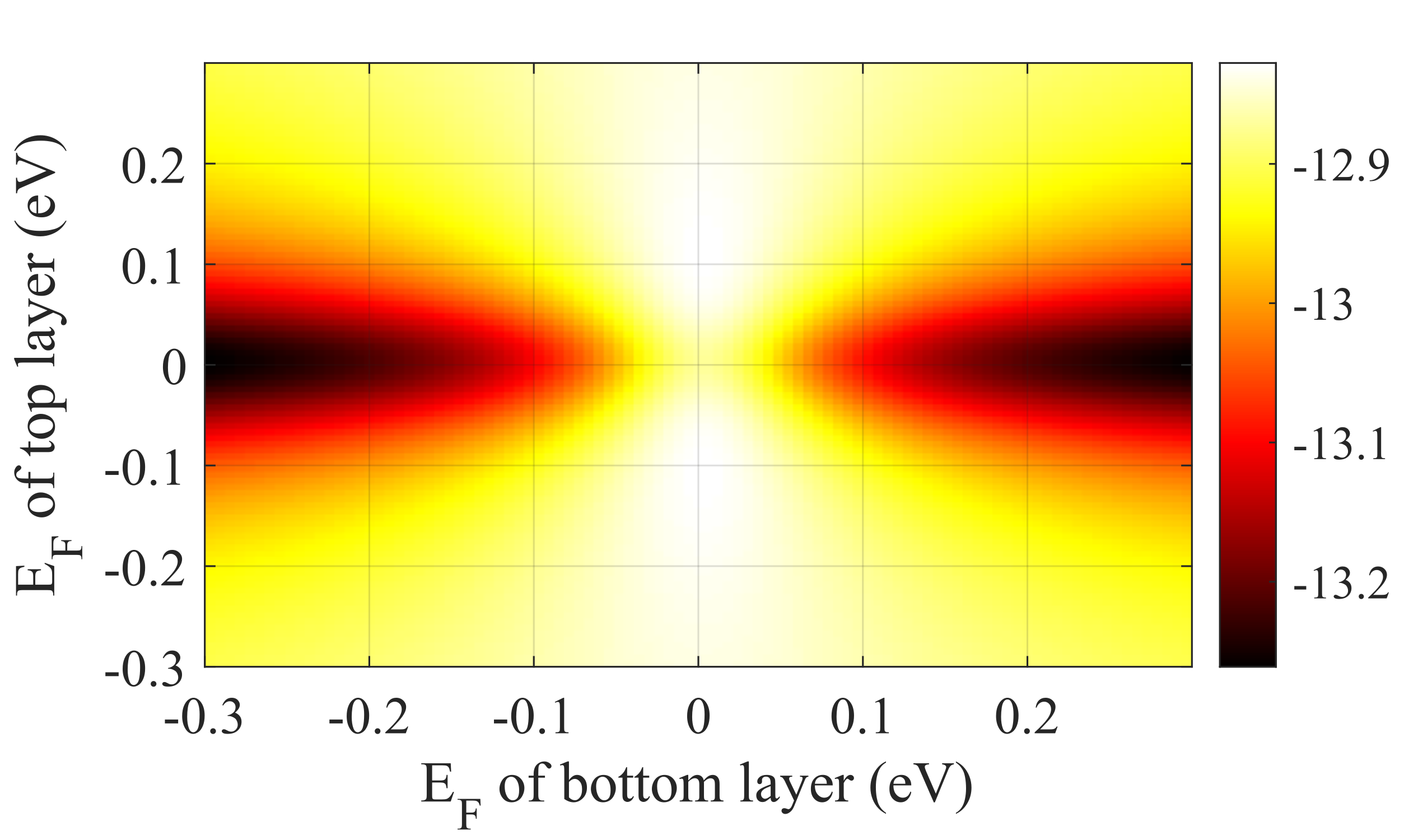}
\caption{The CP potential (in MHz) calculated versus the Fermi energies $E_{F}$ of the top and bottom layer for a \textsuperscript{87}Rb atom in the 30S state. Here, we assume that an hBN slab is sandwiched between graphene sheets. Surprisingly, the potential is not the weakest when both layers are undoped. The parameters are: T = $\SI{300}{K}$, $z_{0} = \SI{2}{\mu}$m and $d = \SI{10}{nm}$.}
\label{fig:30S_E_F_vs_E_F}
\end{figure}

\begin{figure}[ht]
\includegraphics[width = \linewidth]{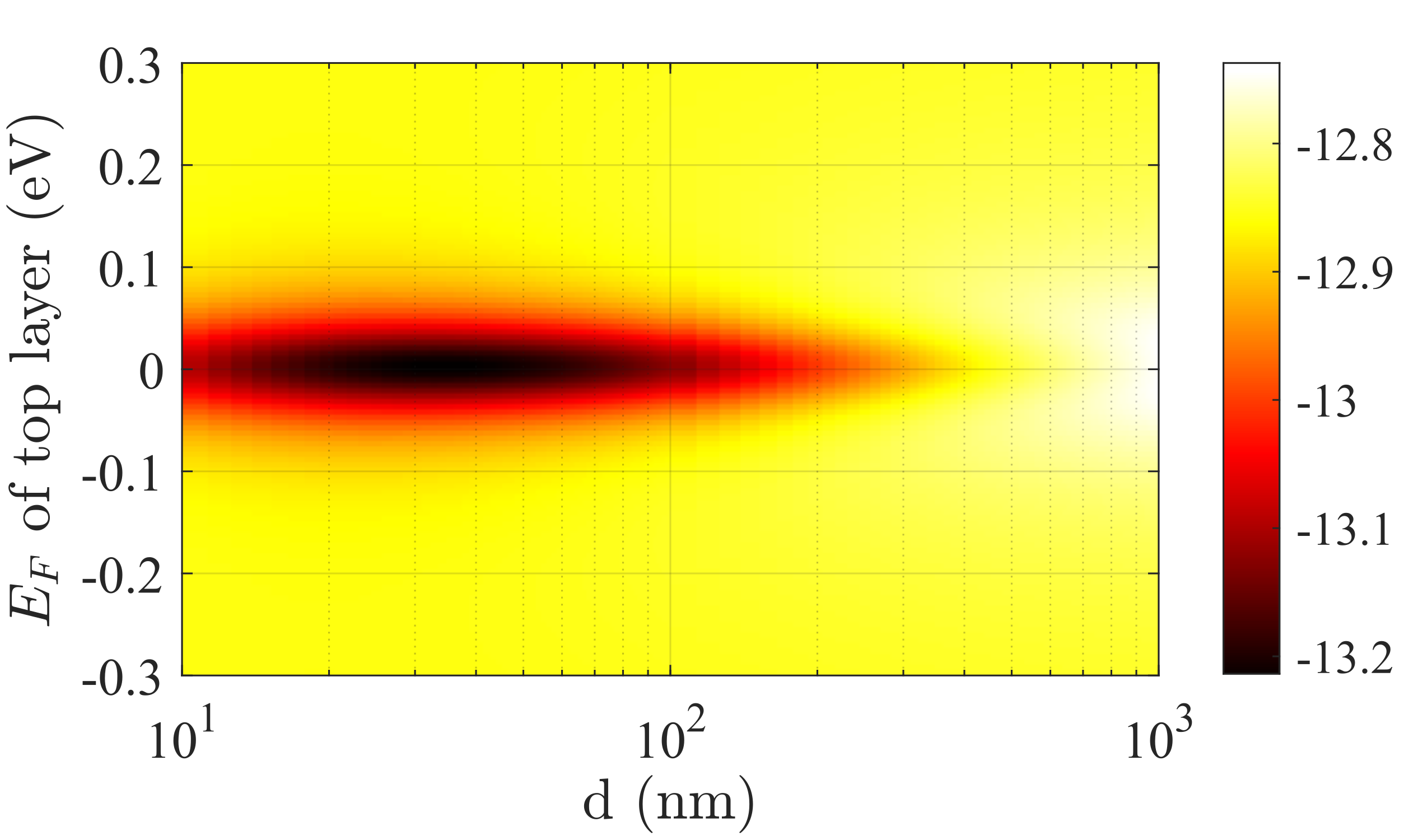}
\caption{The CP potential (in MHz) calculated (and plotted on a log-linear scale ) versus the spacing $d$ between graphene layers and the Fermi energy $E_{F}$ of the top layer for a \textsuperscript{87}Rb atom in the 30S state when the Fermi energy of the bottom layer $= \SI{0}{eV}$. For $d$ below approximately $\SI{400}{nm}$, the potential is the most attractive when $E_{F} = 0$ then gets weaker as $E_{F}$ moves away from $0$ symmetrically. Thereafter, the behavior starts to become similar to that of a single layer of graphene. The parameters are: T = $\SI{300}{K}$, $z_{0} = \SI{2}{\mu}$m.}
\label{fig:30S_d_vs_E_F}
\end{figure}

Now let us consider how the CP potential changes when we simultaneously vary the Fermi levels of the top and bottom graphene layers. As a background, for a ground-state atom near two layers of graphene, the CP potential generally becomes more attractive as graphene sheets are electrically doped \cite{Ribeiro_Scheel_2013}. Figure \ref{fig:30S_E_F_vs_E_F} shows the CP potential (in MHz) versus the Fermi energies $E_{F}$ of the top and bottom layer of the graphene-hBN-graphene structure for a \textsuperscript{87}Rb atom in the 30S state at T = $\SI{300}{K}$, taking $z_{0} = \SI{2}{\mu}$m and $d = \SI{10}{nm}$. We can see that in the middle of the colour map where the Fermi energies of both layers are zero, the potential is still stronger than in most regions of the parameter plane. Moreover, it is very surprising that the potential is the most attractive when $E_{F}$ of the top layer is zero. This cannot simply be explained by considering the conductivity alone; in addition the non-resonant and resonant terms are enhanced differently when the Fermi energies are varied.\\
\\
Lastly, let us investigate how the CP potential depends on the spacing $d$ and on the Fermi energy of the top layer when the Fermi energy of the bottom layer is fixed. Figure \ref{fig:30S_d_vs_E_F} shows the CP potential (in MHz) of the graphene-vacuum-graphene structure calculated versus the spacing $d$ between the graphene layers (in a logarithmic scale) and the Fermi energy $E_{F}$ of the top layer when the Fermi energy of the bottom layer is set to $\SI{0}{eV}$ at $T$ = $\SI{300}{K}$ and $z_{0} = \SI{2}{\mu}$m. We may divide the behavior of the CP potential landscape into three regimes: the first is the SLG-like regime, in which the separation between the two graphene layers is large. In this regime, the potential becomes more negative when the Fermi energy is increased. The second regime is when $d \sim 400$-$\SI{700}{nm}$; the potential then barely changes with Fermi energy. The third regime is when $d < \SI{400}{nm}$. In this regime, the potential is the most attractive when $E_{F} = \SI{0}{eV}$.

\section{Conclusions}
\label{sec:conclusion}
In summary, we have calculated and analyzed the CP potential of a rubidium Rydberg atom positioned near single-layer and double-layer graphene. These calculations used a Green’s-function method in the framework of macroscopic quantum electrodynamics. The optical conductivity of the layer(s) was modeled by the local Kubo equation, including non-local effects. Together, the atomic electric polarizability, the dipole matrix elements of the atom, and the electromagnetic reflection coefficients of the surface layer(s), determine the CP potential. Since Rydberg atoms have higher electric polarizabilities than, and their dipole matrix elements between adjacent atomic states exceed, those of ground-state atoms, their CP potential interaction with graphene-based multilayers is complex.\\
\\
We have shown that, at $T = \SI{300}{K}$ in the non-retarded regime, the CP potential of monolayer graphene is weaker than that for a ${1}$-$\mu$m-thick gold sheet for low-$n$ states, but the CP potential of graphene becomes increasingly attractive when $n > 20$. Regarding the different models of graphene's conductivity, the near-surface potential calculated using the local conductivity is slightly more attractive than that calculated from the full non-local conductivity models for typical doping levels ($E_{F} = \SI{0.1}{eV}$) and $T = \SI{10}{K}$. In general, in the non-retarded limit, the CP potential is determined by both the monotonic attractive non-resonant potential and the evanescent-wave resonant potential. In the retarded limit, the CP potential is dominated by the resonant potential alone and spatially oscillates with a periodicity that approximately equals the half wavelength of the nearest downward transition at low temperature and the nearest upward transition at high temperature. The spatial oscillations in the CP potential start to occur when the atom-surface separation begins to exceed these wavelengths. Thermal effects come into play when the thermal energy is resonant with the atomic transition energies. In the non-retarded, spectroscopic low-temperature limit, the atomic transitions are dominated by downward transitions, which give rise to an attractive resonant potential. In contrast, in the spectroscopic high-temperature limit, there are enough thermal photons to stimulate upward transitions, resulting in a repulsive potential. The spectroscopic high-temperature limit can be easily realized by increasing the principal quantum number even below room temperature. Doping graphene generally results in a more attractive CP potential at high temperatures.\\
\\
Finally, we have investigated heterostructures containing two graphene sheets with varying inter-layer separation and Fermi energies. We found that the effects of changing the spacing and Fermi energies are inter-related. When the spacing is small ($d \ll \SI{1}{\mu}$m), doping the top layer either positively or negatively weakens the CP potential. By contrast, when the spacing is large the behavior approaches that of single-layer graphene.\\
\\
Possible future work could be done on multiple-layer structures comprising graphene and other 2D materials. The complex interference of electromagnetic waves within and near such structures could greatly affect the resonant CP potential. Studying the many-body effects of Rydberg atoms near graphene-based heterostructures might also be interesting since parameters such as the layer chemical potentials, inter-layer spacing, and/or the number and type of 2D layers can be altered to manipulate the interaction with, and behavior of, nearby trapped atoms.\\
\\
Acknowledgements: This work is supported by the EPSRC through Grant No. EP/R04340X/1 via the QuantERA project “ERyQSenS”.

\appendix
\section{Green's tensor}
\label{appendix:Green's tensor}
% The equal-position scattering Green's tensor is given by \cite{buhmann_ii, amorim_2017}
% \begin{multline}
% \mathbf{G}^{(\mathrm{s})}(\mathbf{r}_{0}, \mathbf{r}_{0}, \omega) = \frac{i}{8\pi}\int_{0}^{\infty}\mathrm{d}k^{\parallel}\frac{k^{\parallel}}{k^{\perp}}\mathrm{e}^{2ik^{\perp}z_{0}}\\
% \times \Big[\mathbf{M}_{\alpha}r_{s}^{(\mathrm{s})}(k^{\parallel}, \omega)
% + \frac{c^{2}}{\omega^{2}}
% \mathbf{M}_{\beta}r_{p}^{(\mathrm{s})}(k^{\parallel}, \omega)\Big],
% \label{eq:Green_for_electric_dipole}
% \end{multline}
% where $z_{0}$ is the distance between the surface and the center of the atom and
% \begin{equation}
% k^{\perp} = \Big(\mu_{1}\epsilon_{1}\frac{\omega^{2}}{c^{2}} - k^{\parallel2}\Big)^{1/2},
% \label{eq:k_y1}
% \end{equation}
% with $k^{\parallel2} = k_{x}^{2} + k_{y}^{2}$. \todo{!!}
% The tensors $\mathbf{M}_{\alpha}$ and $\mathbf{M}_{\beta}$ in Eq. \eqref{eq:Green_for_electric_dipole} are given by

% \begin{align}
% \mathbf{M}_{\alpha} = & 
% \begin{pmatrix}
% 1 & 0 & 0\\
% 0 & 1 & 0\\
% 0 & 0 & 0
% \end{pmatrix},\label{eq:M_alpha}\\
% \mathbf{M}_{\beta} = & 
% \begin{pmatrix}
% -k^{\perp2} & 0 & 0\\
% 0 & -k^{\perp2} & 0\\
% 0 & 0 & 2k^{\parallel2}
% \end{pmatrix}.
% \label{eq:M_beta}
% \end{align}
% Note that the forms of $\mathbf{M}_{\alpha}$ and $\mathbf{M}_{\beta}$ depend on the coordinate system used and that Equation \eqref{eq:Green_for_electric_dipole} only describes an electromagnetic field with a \emph{real} frequency, created by a radiating \emph{electric} dipole.\\
An electric field created by a radiating electric dipole can be described by a classical Green's tensor whose form for planar multilayered systems is well-known.
When analysing the resonant part of the CP potential, it is useful to split the equal-position scattering Green's tensor into evanescent-wave and propagating-wave components as follows \cite{buhmann_ii, amorim_2017}:
\begin{equation}
\mathbf{G}^{(\mathrm{s})}(\mathbf{r}_{0}, \mathbf{r}_{0}, \omega) = \mathbf{G}^{(\mathrm{s})}_{\mathrm{evan}}(\mathbf{r}_{0}, \mathbf{r}_{0}, \omega) + \mathbf{G}^{(\mathrm{s})}_{\mathrm{prop}}(\mathbf{r}_{0}, \mathbf{r}_{0}, \omega).
\label{eq:G_two_components}
\end{equation}
The evanescent-wave component takes the form
\begin{multline}
\mathbf{G}^{(\mathrm{s})}_{\mathrm{evan}}(\mathbf{r}_{0}, \mathbf{r}_{0}, \omega) = \frac{1}{8\pi}\int_{0}^{\infty}\mathrm{d}\kappa^{\perp}\mathrm{e}^{-2\kappa^{\perp}z_{0}}\\
\times \Big[\mathbf{M}_{\alpha}r_{s}^{(\mathrm{s})}(k^{\parallel}, \omega)
+ \frac{c^{2}}{\omega^{2}}
\mathbf{M}_{\beta}r_{p}^{(\mathrm{s})}(k^{\parallel}, \omega)\Big],
\label{eq:Green_evan}
\end{multline}
and the propagating-wave component takes the form
\begin{multline}
\mathbf{G}^{(\mathrm{s})}_{\mathrm{prop}}(\mathbf{r}_{0}, \mathbf{r}_{0}, \omega) = \frac{i}{8\pi}\int_{0}^{\omega/c}\mathrm{d}k^{\perp}\mathrm{e}^{2ik^{\perp}z_{0}}\\
\times \Big[\mathbf{M}_{\alpha}r_{s}^{(\mathrm{s})}(k^{\parallel}, \omega)
+ \frac{c^{2}}{\omega^{2}}
\mathbf{M}_{\beta}r_{p}^{(\mathrm{s})}(k^{\parallel}, \omega)\Big],
\label{eq:Green_prop}
\end{multline}
where $z_{0}$ is the shortest distance between the surface and the center of the atom, $r_{s}^{(\mathrm{s})}$ and $r_{p}^{(\mathrm{s})}$ are the Fresnel reflection coefficients for the $s$- and $p$-polarized waves, respectively, and
\begin{equation}
\kappa^{\perp} = \Big(k^{\parallel2} - \mu_{1}\epsilon_{1}\frac{\omega^{2}}{c^{2}}\Big)^{1/2},\\
k^{\perp} = \Big(\mu_{1}\epsilon_{1}\frac{\omega^{2}}{c^{2}} - k^{\parallel2}\Big)^{1/2}
\label{eq:k_y1}
\end{equation}
with $k^{\parallel2} = k_{x}^{2} + k_{y}^{2}$. \\
The tensors $\mathbf{M}_{\alpha}$ and $\mathbf{M}_{\beta}$ in Eqs. \eqref{eq:Green_evan} and \eqref{eq:Green_prop} are given by

\begin{align}
\mathbf{M}_{\alpha} = & 
\begin{pmatrix}
1 & 0 & 0\\
0 & 1 & 0\\
0 & 0 & 0
\end{pmatrix},
\label{eq:M_alpha}\\
\mathbf{M}_{\beta} = & \begin{cases}
\begin{pmatrix}
\kappa^{\perp2} & 0 & 0\\
0 & \kappa^{\perp2} & 0\\
0 & 0 & 2k^{\parallel2}
\end{pmatrix}
\text{for $\mathbf{G}^{(\mathrm{s})}_{\mathrm{evan}}$},\\
\\
\begin{pmatrix}
-k^{\perp2} & 0 & 0\\
0 & -k^{\perp2} & 0\\
0 & 0 & 2k^{\parallel2}
\end{pmatrix}
\text{for $\mathbf{G}^{(\mathrm{s})}_{\mathrm{prop}}$}.
\end{cases}
\label{eq:M_beta}
\end{align}

\section{Graphene's Optical Properties}
\label{appendix:Graphene's Optical Properties}
\subsection{Conductivity models}
In this section, we present two models for the optical conductivity of monolayer graphene: (i) based on local Kubo conductivity, $\sigma^{\mathrm{Kubo}}(\omega)$, which depends only on the angular frequencies, $\omega$, of the incident electromagnetic radiations and ignores spatial dispersion in the graphene surface \cite{Goncalves2016, stauber_peres_geim_2008, hanson_2013} and (ii) using the full non-local conductivity, $\sigma^{\mathrm{fnl}}(q, \omega)$, which is derived from the Lindhard polarisation function in random-phase (RPA) and relaxation-time (RT) approximations \cite{Linhard_1954,Wunsch_2006,Hwang_Sarma_2007,Goncalves2016}. The latter takes into account spatial dispersion by non-local causes when considering the interactions of incident photons, surface plasmons with wavenumbers, $q$, and graphene's electrons.\\

The Kubo conductivity can be expressed as the sum of two contributions: $\sigma_{\mathrm{intra}}(\omega)$, which arises from intraband transition processes of the electrons and $\sigma_{\mathrm{inter}}(\omega)$, which describes transitions between the conduction and valence bands; both terms may be written as follows:

\begin{equation}
\sigma_{\mathrm{intra}}(\omega) = \frac{\sigma_{0}}{\pi}\frac{4}{\hbar\gamma -i\hbar\omega} \bigg[E_{F} + 2k_{B}T \ln \bigg( 1 + e^{-E_{F}/k_{B}T} \bigg) \bigg], 
\label{eq:conductivity_intra}
\end{equation}

\begin{equation}
\sigma_{\mathrm{inter}}(\omega) = \sigma_{0}\bigg[G(\hbar\omega/2) + i\frac{4\hbar\omega}{\pi}\int_{0}^{\infty}dE\frac{G(E) - G(\hbar\omega/2)}{(\hbar\omega)^{2} - 4E^{2}}\bigg],
\label{eq:conductivity_inter}
\end{equation}
in which

\begin{equation}
G(X) = \frac{\sinh\bigg(\frac{\displaystyle X}{\displaystyle k_{B}T}\bigg)}{\cosh\bigg(\frac{\displaystyle E_{F}}{\displaystyle k_{B}T}\bigg) + \cosh\bigg(\frac{\displaystyle X}{\displaystyle k_{B}T}\bigg)},
\label{eq:G(x)_function}
\end{equation}
where $\sigma_{0} = e^{2}/(4\hbar)$ is the universal alternating-current conductivity of graphene, $\gamma$ is the electron relaxation rate in graphene, $E_{F}$ is the Fermi energy and $T$ is the temperature of the graphene layer.\\

\begin{figure}[ht]
\includegraphics[width = 0.6\linewidth]{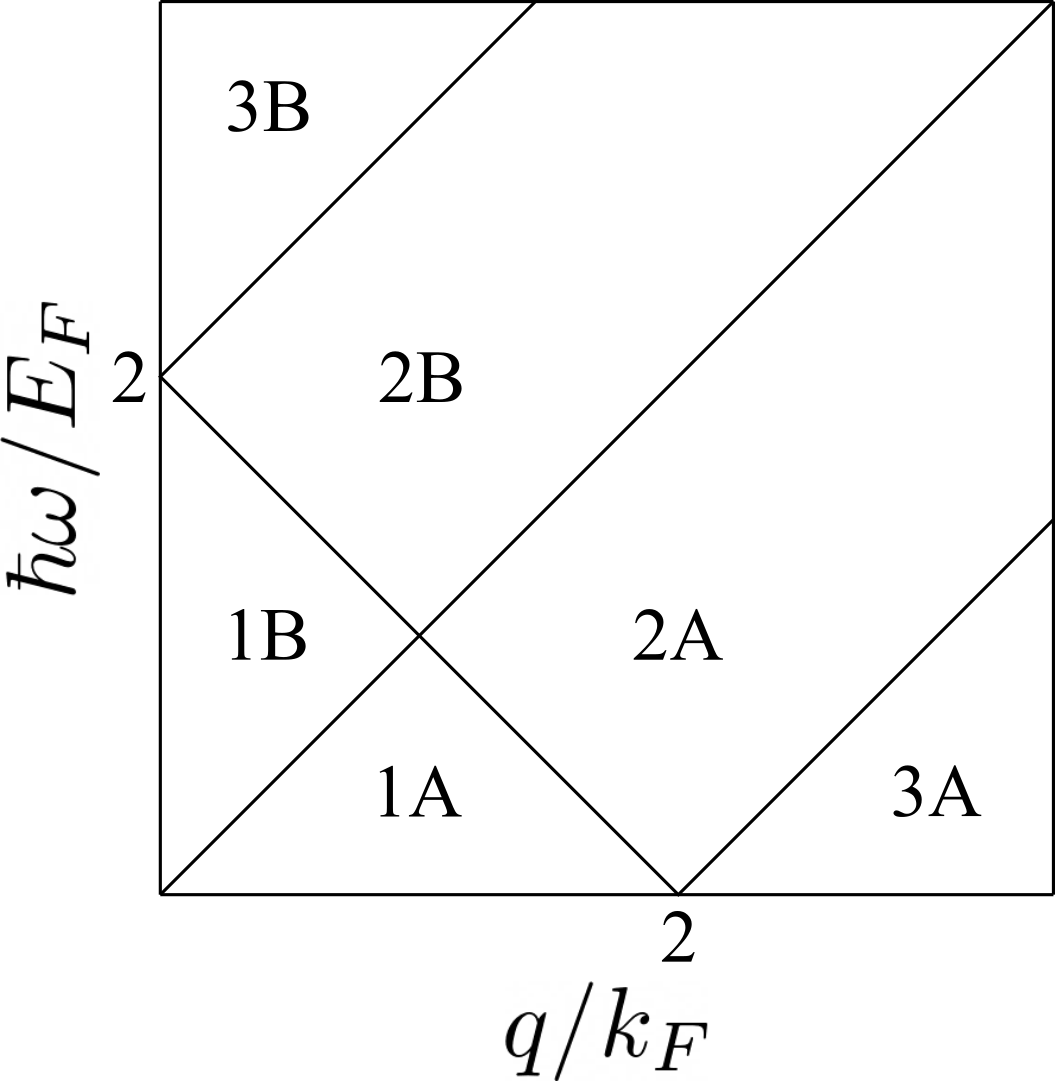}
\caption{Schematic diagram showing six regions in the $(q, \hbar\omega)$-plane of graphene surface-plasmons (GSP), which are distinguished by the coupling between the plasmons and excitations of electron-hole pairs in graphene.}
\label{fig:qomega_map}
\end{figure}

As for the other model--the full non-local conductivity, we start by providing the 2D polarizability in the RPA, $P(q, \omega) = P_{re}(x, y) + iP_{im}(x,y)$, in the $T = \SI{0}{K}$ limit, where $P_{re}(x, y)$ and $P_{im}(x,y)$ are the real and imaginary parts, respectively, in six regions of the $(q, \hbar\omega)$-space as depicted in Fig. \ref{fig:qomega_map} as follows: \\

$\bullet$ region 1A
\begin{align} 
P_{re}(x, y) =& -2t_{1},\\
P_{im}(x, y) =& \phantom{i}\frac{1}{4}t_{1}t_{3}[C_{h}(t_{5}) - C_{h}(t_{4})],
\label{eq:1A}
\end{align}

$\bullet$ region 2A
\begin{align} 
P_{re}(x, y) =& -2t_{1} + \frac{1}{4}t_{1}t_{3}C(t_{5}),\\
P_{im}(x, y) =& -\frac{1}{4}t_{1}t_{3}C_{h}(t_{4}),
\label{eq:2A}
\end{align}

$\bullet$ region 3A
\begin{align} 
P_{re}(x, y) =& -2t_{1} + \frac{1}{4}t_{1}t_{3}[C(t_{4}) - C(t_{6})],\\
P_{im}(x, y) =& \phantom{i}0,
\label{eq:3A}
\end{align}

$\bullet$ region 1B
\begin{align} 
P_{re}(x, y) =& -2t_{1} + \frac{1}{4}t_{1}t_{2}[C_{h}(t_{4}) - C_{h}(t_{5})],\\
P_{im}(x, y) =& \phantom{i}0,
\label{eq:1B}
\end{align}

$\bullet$ region 2B
\begin{align} 
P_{re}(x, y) =& -2t_{1} + \frac{1}{4}t_{1}t_{2}C_{h}(t_{4}),\\
P_{im}(x, y) =& \phantom{i}\frac{1}{4}t_{1}t_{2}C(t_{5}),
\label{eq:2B}
\end{align}

$\bullet$ region 3B
\begin{align} 
P_{re}(x, y) =& -2t_{1} + \frac{1}{4}t_{1}t_{2}[C_{h}(t_{4}) - C_{h}(t_{6})],\\
P_{im}(x, y) =& -\frac{\pi}{4}t_{1}t_{2},
\label{eq:3B}
\end{align}
where $x = q/k_{F}$, $y = \hbar\omega/E_{F}$, $t_{1} = {k_{F}}/{\pi\hbar v_{F}}$, $t_{2} = x^{2}/\sqrt{y^{2} - x^{2}}$, $t_{3} = x^{2}/\sqrt{x^{2} - y^{2}}$, $t_{4} = (2+y)/x$, $t_{5} = (2-y)/x$, $t_{6} = (y-2)/x$. Here $v_{F}$ and $k_{F}$ are Fermi velocity and wavenumbers, respectively. The auxiliary functions are defined as follows

\begin{align} 
C_{h}(a) =& a\sqrt{a^{2} - 1} - \mathrm{arccosh}(a),\\
C(a) =& a\sqrt{1 - a^{2}} - \mathrm{arccos}(a).
\label{eq:Ch_and_C}
\end{align}

The 2D polarizability described above only takes into account intrinsic mechanisms for the decay of GSP into electron-hole pairs. To include extrinsic processes such as collisions with lattice defects or impurity scattering, we extend the previous model by also including scattering events within the relaxation-time approximation, which allows us to express the 2D polarizability in the RPA-RT approximation as follows:

\begin{equation}
P_{\gamma}(q,\omega)= \frac{(1+i\gamma/\omega)P(q,\omega+\mathrm{i\gamma})}{(1+i\gamma/\omega)\cdot P(q,\omega+\mathrm{i\gamma})/P(q,0)}.
\label{eq:P_gamma}
\end{equation}

The RPA-RT dielectric function can be written in terms of the 2D polarisation function as

\begin{equation}
\epsilon^{\mathrm{RPA-RT}}(q,\omega) = \epsilon_{r} - v_{q}P_{\gamma}(q,\omega),
\label{eq:RPA-RT epsilon}
\end{equation}
where $\epsilon_{r}$ is the relative permittivity of the medium in which the graphene layer is embedded and $v_{q} = e^{2}/2\epsilon_{0}q$ is the Fourier transform of the Coulomb interaction. Additionally, the longitudinal conductivity can also be written in terms of the Lindhard polarizability: 

\begin{equation}
\sigma^{\mathrm{fnl}}(q,\omega) = ie^{2}\frac{\omega}{q}P_{\gamma}(q,\omega).
\label{eq:full non-local conductivity}
\end{equation}
Note that it is this quantity--the conductivity--that will be inserted into our transfer-matrix calculations of the reflection coefficients of graphene, when we calculate the CP potential.\\

\begin{figure}[ht]
\includegraphics[width = \linewidth]{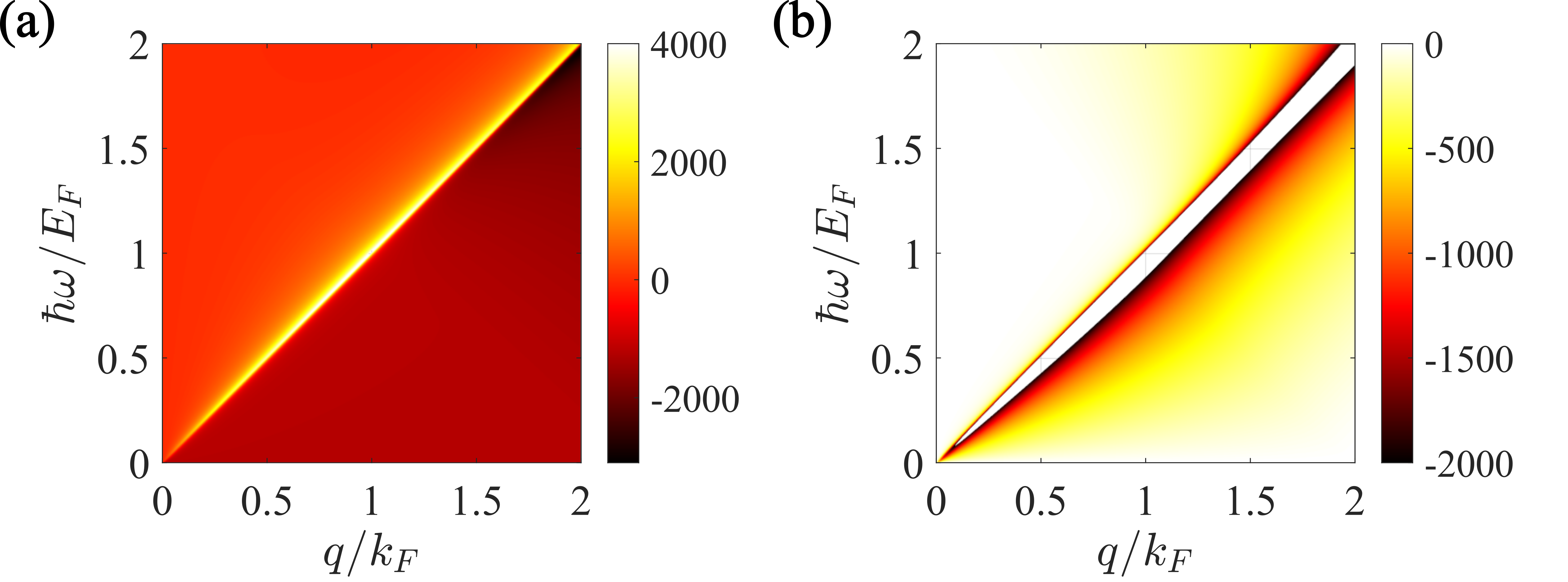}
\caption{Real (a) and imaginary (b) parts of the 2D polarizability of graphene, $P_{\gamma}(q,\omega)\times h$, plotted in four regions--1A, 2A, 1B and 2B--of the $(q, \hbar \omega)$ plane. The parameters are $\gamma = \SI{4}{THz}$, $E_{F} = \SI{0.1}{eV}$.}
\label{fig:P_gamma_map}
\end{figure}

Fig. \ref{fig:P_gamma_map} shows the real and imaginary parts of the 2D polarizability of graphene, calculated using Eq. \eqref{eq:P_gamma}. Note that the imaginary part in these regions is always negative.\\

% Now that we have the expression of the conductivity for both models, let us investigate the differences between them as we vary the frequency, $\omega$, and the wave number, $q$.
% Fig. \ref{fig:New_conductivity} shows the Kubo (solid curve) and the full non-local (dashed curves) conductivity for different values of $q$ as a function of $\omega$. We can see that the full non-local conductivity deviates slightly from the Kubo conductivity as we raise $q$ to $10^{-6}k_{F}$. The position of the peak shifts to higher $\omega$ while the height of the peak drops as $q$ increases. 

% \begin{figure}[ht]
% \includegraphics[width = \linewidth]{New_conductivity.png}
% \caption{\todo{change the last three to dashed curves}}
% \label{fig:New_conductivity}
% \end{figure}

% \subsection{Graphene surface plasmons}
% \begin{figure}[ht]
% \includegraphics[width = \linewidth]{LossFunctionPlot.png}
% \caption{Free-standing graphene's loss function computed using two different material properties: (a) $L(q,\omega) =-\Im{\epsilon_{\mathrm{RPA-RT}}^{-1}(q,\omega)}$ and (b) $L(q,\omega) =\Im{r_{p}(q,\omega)}$ for Fermi energies $E_{F} = \SI{0.1}{eV}$ (left) and $E_{F} = \SI{0.2}{eV}$ (right). The intensity plots are presented in a $\log_{10}[1+L(q,\omega)]$ scale with a hot colour scheme. The blue dashed lines separate regions A, 1B and 2B (see Fig. \ref{fig:qomega_map}). The GSP band (hot colours) mainly exists in region 1B and fades away when it enters region 2B. The FWHM of the GSP band is inversely proportional to $E_{F}$. Parameter: $\gamma = \SI{4}{THz}$.}
% \label{fig:LossFunctionPlot}
% \end{figure}

\section{Permittivity of metals}
The permittivity of gold at radiation frequency $\omega$ is described by the Drude model \cite{ashcroft_mermin_1976}

\begin{equation}
\epsilon_{\mathrm{metal}}(\omega) = 1 - \frac{\omega_{\mathrm{p}}^{2}}{\omega^{2} + i\Gamma_{D}\omega},
\label{eq:Drude_model}
\end{equation}
where, for gold, $\omega_{\mathrm{p}} = \SI{1.35e16}{rad/s}$ is the plasma frequency and $\Gamma_{D} = \SI{17.13}{THz}$ is the electron scattering rate \cite{Zeman_George_1987}.

\newpage
\bibliographystyle{unsrt}
\bibliographystyle{apsrev4-2}
\bibliography{NewRef}

\end{document}